\documentclass{gGAF2e}
\pdfoutput=1
\usepackage[utf8]{inputenc}
\usepackage[T1]{fontenc}
\usepackage{amsmath}
\usepackage{amssymb}
\usepackage{graphicx}
\usepackage{bm}
\usepackage{indentfirst}
\usepackage{subfigure}
\usepackage{caption}
\usepackage{pdfpages}
\usepackage{multirow}
\usepackage{multicol,graphicx}
\usepackage{mathtools}

\newcommand{\araa         }{Annu.~Rev.~Astron.~Astrophys.}
\newcommand{\apj          }{Astrophys.~J.}

\newcommand{\aap          }{Astron.~Astrophys.}

\newcommand{\eps          }{Earth Planets Space}

\newcommand{\gafd         }{Geophys. Astrophys. Fluid Dyn.}
\newcommand{\gji          }{Geophys. J. Int.}
\newcommand{\grl          }{Geophys. Res. Lett.}

\newcommand{\jfm          }{J. Fluid Mech.}
\newcommand{\jgg          }{J. Geomag. Geoelec.}

\newcommand{\jgrb         }{J. Geophys. Res. B Solid Earth Planets}
\newcommand{\jpc          }{J. Phys. Conf.}

\newcommand{\mnras        }{Mon. Not. R. Astron. Soc.}
\newcommand{\nat          }{Nature}
\newcommand{\natg         }{Nat. Geosci.}

\newcommand{\pepi         }{Phys. Earth Planet. Inter.}

\newcommand{\prsla        }{Proc. Roy. Soc. Lond. A}

\newcommand{\rpr          }{Rep. Progr. Phys.}

\newcommand{\rmp          }{Rev. Mod. Phys.}
\newcommand{\rsos         }{Royal Soc. Open Sci.}
\newcommand{\sa           }{Sci. Adv.}

\newcommand{\rmd}{{\mathrm d}}

\renewcommand{\Lambda}{{\varLambda}}
\renewcommand{\Omega}{{\varOmega}}
\renewcommand{\Theta}{{\varTheta}}

\newcommand*{\bse}{\begin{subequations}}
\newcommand*{\ese}{\end{subequations}}

\begin{document}

\title{Magnetohydrodynamics of stably stratified regions in planets and stars}

\author{J.Philidet\textsuperscript{${\dagger}$}, C.Gissinger\textsuperscript{${\ddagger}$}, F.Ligni\`eres\textsuperscript{${\S}$} and L.Petitdemange\thanks{$^\ast$Corresponding author. Email: ludovic@lra.ens.fr}$^\ast$ \textsuperscript{${\P}$} \\
		\vspace{6pt}
		${\dagger}$ LESIA, Observatoire de Paris, PSL Research University, CNRS, Universit\'e Pierre et Marie Curie, Universit\'e Paris Diderot, 92195 Meudon, France \\
		${\ddagger}$ Laboratoire de Physique de l’Ecole Normale Sup\'erieure, ENS, Universit\'e PSL, CNRS, 24 rue Lhomond, 75005 Paris, France \\
		${\S}$ IRAP, Université de Toulouse, CNRS, CNES, UPS, F-31400 Toulouse, France \\
		${\P}$ LERMA, Observatoire de Paris, PSL Research University, CNRS, Sorbonne Universit\'es, UPMC Univ. Paris 06, Ecole Normale Sup\'erieure, 75005 Paris, France
		}

\markboth{\rm {J.~PHILIDET ET AL.}}{\rm {GEOPHYSICAL $\&$ ASTROPHYSICAL FLUID DYNAMICS}}

\maketitle

\begin{abstract}
Stably stratified layers are present in stellar interiors (radiative zones) as well as planetary interiors - recent observations and theoretical studies of the Earth's magnetic field seem to indicate the presence of a thin, stably stratified layer at the top of the liquid outer core. We present direct numerical simulations of this region, which is modelled as an axisymmetric spherical Couette flow for a stably stratified fluid embedded in a dipolar magnetic field. For strong magnetic fields, a super-rotating shear layer, rotating nearly $30\%$ faster than the imposed rotation rate difference between the inner convective dynamo region and the outer boundary, is generated in the stably stratified region. In the Earth context, and contrary to what was previously believed, we show that this super-rotation may extend towards the Earth magnetostrophic regime if the density stratification is sufficiently large. The corresponding differential rotation triggers magnetohydrodynamic instabilities and waves in the stratified region, which feature growth rates comparable to the observed timescale for geomagnetic secular variations and jerks. In the stellar context, we perform a linear analysis which shows that similar instabilities are likely to arise, and we argue that it may play a role in explaining the observed magnetic dichotomy among intermediate-mass stars.
\end{abstract}


\section{Introduction}\label{sec:intro}

Understanding the origin of the diversity of stellar magnetic fields is a research topic in which important advances have been recently made and much more are expected soon. This topic is pushed by new observations of ground-based instruments, in particular spectropolarimeters, as well as observations from space (with \textit{CoRoT} \citealt{baglin06} and Kepler \citealt{borucki10} for instance). It also benefits from the results of simulations which, thanks to the numerical resources now available and the progress of astrophysical modelling, allow us to study the complex physics of stellar interiors taking into account many effects such as rotation, magnetism, turbulence\ldots While it is accepted that magnetism can play a major role in most stages of a star life, its effects are mainly ignored in stellar evolution codes. Taking into account these magnetic effects is one of the major current issues in astrophysics. 

Stellar magnetic fields can be classified into two broad categories. In the first category, magnetic fields are generated and maintained by convective motions, through what is referred to as dynamo effect. There are countless different mechanisms through which dynamo can give rise either to stationary fields, large-scale dipoles like those observed for rapidly rotating low-mass, M type stars, or to time-dependent fields. The periodic variation of the solar field has been observed since Galileo through the counting of solar dark spots. For these stars having masses less than or equal to the solar mass, dynamo mechanisms seem to be very efficient in their thick convective envelope, an idea supported by many studies on planetary magnetism \citep{christensenHR09, robertsK13, petitdemange18}. In the second category, on the other hand, massive and intermediate mass stars have a thick radiative envelope above a small convective core, and harbor strong dipolar fields with a stable configuration, commonly referred to as ``fossil fields'', because they do not show time variations and it is believed to originate from an earlier stage of the star's life. Unlike dynamo-driven fields, the origin of magnetism in stars possessing a stably stratified outer layer is poorly explored by theoretical studies, and understanding it remains a major challenge \citep[see][for a review]{braithwaiteS17}.

Thanks to the improved sensitivity of a new generation of spectropolarimeters, the existence of a dichotomy in the magnetic field amplitudes for intermediate mass stars of the main-sequence is now clear \citep{lignieres14}. Indeed, the observations show a large gap between strong fields (fossil fields with amplitudes above $300$ G) and fields of ultra-weak, sub-gauss magnitude. This gap, referred to as the ``magnetic desert'', stretches over two orders of magnitude. It has been proposed in \citet{auriere07} that the observed magnetic dichotomy is a consequence of a bifurcation between stable and unstable large scale magnetic configurations in differentially rotating radiative envelopes. While strong magnetic fields can suppress differential rotation \citep{moss92, spruit99, jouveGL15}, weaker magnetic fields might not inhibit differential rotation before it leads to MHD instabilities that will affect the field large scale geometry. A better understanding of the stability of magnetic fields in such stably stratified layers is required to investigate this scenario in detail \citep{jouveGL15, gaurat15}.

In addition, differential rotation and magnetism appear as intimately linked phenomena. Differential rotation, by itself or by developing hydrodynamic instabilities, can generate dynamo-generated magnetic fields, and this magnetism can inhibit these instabilities or even greatly reduce the differential rotation by transporting angular momentum outwards. This last process could explain the magnitude of the differential rotation observed by asteroseismology for different stars \citep{deheuvels14}.

MHD phenomena are crucial to the understanding of planetary interiors as well. Earth's magnetic field is known to be generated by dynamo effects taking place in its liquid outer core. The study of this region in the scope of MHD may therefore give some insight into the variations of Earth's magnetic field. Those take place on time scales ranging from less than a year to millions of years. In particular, the variations of the length of day on timescales ranging from 1 to 100 years are believed to be caused by geomagnetic fluctuations featuring these typical periods, so called secular variations (SV). However, while being observed for a long time and monitored with an ever increasing accuracy \citep{chulliat14,chulliat15,finlay16}, geomagnetic variations remain a fascinating problem as several features of the so-called magnetic jerks are still unclear. In particular, it was recently suggested \citep{buffett16} that the presence of a certain type of MHD waves resulting from the interaction between magnetic tension, Archimedes force and Coriolis force in a $140$ km thin stably stratified layer located at the top of Earth's liquid outer core \citep{braginsky93} may account for unexplained secular fluctuations of the geomagnetic dipole field. The presence of this layer at the core-mantle boundary (CMB) is believed to be due to the diffusion of light elements from the mantle to the core driven entirely by pressure gradients \citep{gubbins13}, although it may also be originated in thermal diffusion, as investigated by \citet{pozzo12}, who conducted new calculations of the heat conductivity in the Earth core and found that transport from thermal conduction alone may not explain the actual heat flux in the Earth CMB. Note that the existence of this stably-stratified layer on top of the Earth core is still somewhat controversial \citep[see for instance][]{alexandrakis10, irving18}. However, our model is relevant to the planetary context as a whole, and we choose to apply it to the Earth case in this paper because of the abundance of data available compared to other planets. Furthermore, while a recent study by \citet{mound19} has shown that this stably-stratified region may not have spherical symmetry, we will model it as a global spherical layer in this paper. Finally, note that magnetic jerks may not be related to MHD instabilities, and other possible origins are investigated as well in other studies \citep[see for instance][]{lesur18}.

Among the impacts a magnetic field can have on a flow embedded in it, one is the development of MHD instabilities. Under the conditions existing in a stellar radiative region, three main types of instabilities are likely to occur, which differ by the source of free energy they derive from and the energy form they release. As far as released energy is concerned, it can be gravitational energy for Parker-like instabilities \citep{parker66}, magnetic energy for purely magnetic instabilities (like the Tayler instability \citep{tayler73} or those induced by a gradient of toroidal magnetic field lines \citep{achesonH73}). As for their energy reservoir, several instabilities derive from the free kinetic energy dispensed by differential rotation, including shear instabilities, like the magneto-rotational instability MRI or strato-rotational instability SRI (see \citet{balbusH91,balbusH98,menou04} for the MRI and \citet{shalybkov05} for the SRI). In the simulations presented in this paper, however, we focus on the low Rossby number regime in which an axisymmetric steady solution exists, and these hydrodynamic or MHD instabilities are not taken into account. We postpone the numerical study of three dimensional structures induced by a strong differential rotation or by non-axisymmetric instabilities. We only argue on the possible development of MHD instabilities by means of analytical developments.

In this paper, we therefore perform a direct, global numerical modelling of stably stratified layers, by taking into account both the rotational influence of neighboring layers and the effect of the magnetic field generated in the conducting convective zone, and use this model to describe both the stably stratified layer at the Earth CMB and stellar radiative zones. We present the numerical model in section \ref{sec:model}; in section \ref{sec:hydro} and \ref{sec:superrot}, we discuss the steady-state flow configuration yielded by these simulations. We show the influence of rotation and stratification on the geometry of the flow in purely hydrodynamical models in section \ref{sec:hydro} and we focus on the development of super-rotation induced by the combined influences of magnetism and stratification in section \ref{sec:superrot}. In section \ref{sec:local}, we perform a local, linear stability analysis to show that the combined effects of shear, magnetic field and stratification could yield MHD instabilities able to account for geomagnetic secular variations in the planetary context, or that they could affect the stability of dipolar fields observed for massive stars in the stellar context.

\section{Numerical model}\label{sec:model}

In the present paper, we aim at modelling stably stratified layers in the geophysical and astrophysical contexts. In the planetary context, this corresponds to the $140$ km thin shell located at the Core-Mantle Boundary (CMB), between the fluid, convective dynamo region and the solid mantle. In the stellar context, it corresponds to a radiative zone (more precisely a radiative envelope in the case of a massive or intermediate mass star). In both cases, the region that is being modelled is located between two spheres, and the aspect ratio in these two situations is drastically different (the planetary case is in the thin gap limit, whereas the stellar context is in the thick gap limit). It is important to stress that a 3D modelling of such a flow would be extremely time-consuming when it comes to use parameter values relevant to both regimes. In particular, the Ekman number $E$, which measures the ratio between the viscous force and the Coriolis force, is close to $E \sim 10^{-15}$ in both cases, whereas classical DNS can only reach $E \gtrsim 10^{-7} - 10^{-8}$ \citep{schaefferJNF17}. Furthermore, while classical models are essential to understand the evolution of magnetic fields on the magnetic dissipation time scale (about $10 000$ years), they are less suitable to the study of much shorter timescales (less than a century), such as we are here. For these different reasons, we restrict ourselves to an axisymmetric study of the system. In performing an axisymmetric study, we implicitly preclude non-axisymmetric perturbations from triggering magnetic instabilities, and from inhibiting differential rotation. However, we only consider weak differential rotation in this paper, and these magnetic instabilities tend to be stabilised by rapid global rotation. Consequently, while this limitation must not be overlooked, the conclusions drawn from this study do not constitute unreasonable extrapolation.

The stably stratified layer is therefore modelled as an axisymmetric spherical Couette flow (that is a fluid comprised between two concentric rotating spheres) embedded in a dipolar magnetic field. The inner sphere of radius $r_i$ in which this magnetic field is generated represents the transition to the convective region in both contexts, and has the same conductivity as the modelled fluid. The outer sphere of radius $r_o$, however, is insulating. In the planetary context, magnetic torques acting on the convective (inner) region make it rotate slightly faster than the outer, insulating sphere. Indeed, the stably-stratified layer that we consider in this study could be subjected to a weak differential rotation (low Rossby number limit) as the deeper convectively unstable dynamo region and the solid mantle could rotate with slightly different rates. For instance, \citet{buffett16} argued that MAC waves could transport angular momentum in the stably-stratified layer, and thus influence the length of day. Furthermore, \citet{takehiroS18}, by performing a magnetoconvection study with an upper stably-stratified layer, have shown that zonal flows can penetrate this layer, thus transporting angular momentum. Differential rotation is also of primary importance in the stellar context. We model this by applying a slightly different rotation rate to the two spheres, which will propagate to the fluid in between through no-slip mechanical boundary conditions. The calculations being performed in the frame of reference of the outer sphere, the latter is motionless, whereas the inner sphere has a slow rotation (compared to the global rotation of the frame of reference). Finally, a stable stratification is applied through a temperature difference between the inner and outer spheres.

We consider the Boussinesq approximation, allowing us to neglect variations of the fluid density except in the buoyancy term, leading to the following dimensionless MHD equations:
\bse
\label{eq:full_set}
\begin{align}
\dfrac{\upartial\bm{v}}{\upartial t} \,=\, &\,-\,(\bm{v}{\bm \cdot}\bm{\nabla})\bm{v} + E\Delta \bm{v} - 2\bm{e_z}\times\bm{v} - \bm{\nabla} P \nonumber\\
&\,+ \dfrac{E\Lambda}{Pm}(\bm{\nabla}\times\bm{B})\times\bm{B} + E\widetilde{Ra}\Theta\bm{e_r}\, , \\
\dfrac{\upartial\bm{B}}{\upartial t}\, =\, &\,\dfrac{E}{Pm}\Delta\bm{B} + {\bm \nabla}\times(\bm{v}\times\bm{B})\, , \\
\dfrac{\upartial T}{\upartial t}\, = \,&\,-(\bm{v}{\bm \cdot}\bm{\nabla}) T + \dfrac{E}{Pr}\Delta T\, ,
\end{align}
\ese
%
%
%
where $\bm{v}$, $P$, $\bm{B}$ and $T$ are the dimensionless velocity, pressure, magnetic field and temperature, $\Theta$ is the temperature fluctuation as a function of which we express the density fluctuations in the buoyancy term, $\bm{e_z}$ the unit vector representing the rotation axis and $\bm{e_r}$ the local radial unit vector. Time scales have been normalized by $\Omega_o^{-1}$ (where $\Omega_o$ is the rotation rate of the outer sphere), spatial variables by the radius $r_o$ of the outer sphere, the magnetic field by its amplitude $B_o$ at $r=r_o$ in the equatorial plane and temperatures by the temperature difference $\Delta T$ applied between the two spheres. We introduce the following non-dimensional numbers: the Ekman number $E = \nu / \left(\Omega_o r_o^{2}\right)$ (ratio between viscous and Coriolis forces), the Elsasser number $\Lambda = B_o^{2} / \left(\mu\rho\eta\Omega_o\right)$ (ratio between the Lorentz and Coriolis forces), the magnetic Prandtl number $Pm = \nu / \eta$, the (modified) Rayleigh number $\widetilde{Ra} = \alpha g\Delta T r_o / \left(\nu\Omega_o\right)$ (which measures the amplitude of stratification), the thermal Prandtl number $Pr = \nu / \kappa$ and the Rossby number $Ro = (\Omega_i - \Omega_o) / \Omega_o$. Following traditional notations, we denote the momentum diffusivity of the fluid as $\nu$, its magnetic permeability as $\mu$, its mean density as $\rho$, its magnetic resistivity as $\eta$, its thermal expansion coefficient as $\alpha$, its thermal diffusivity as $\kappa$ and the gravitational acceleration as $g$. In the Boussinesq approximation, the latter is simply expressed as $g = -4\pi G\rho r / 3$, where $G$ is the gravitationnal constant.

\begin{figure*}
 \centering
\begin{tabular}{cc|cc|ccc}
& {\footnotesize $Pr\left(\frac{N}{\Omega}\right)^2 \leq E^{2/3}$}  & & &  \multicolumn{2}{c}{{\footnotesize $Pr\left(\frac{N}{\Omega}\right)^2  > 1$}} &  \\
\rotatebox{90}{{\footnotesize $E=10^{-5}$, $Pr=1$}} &
\subfigure[$Q=10^{-4}$]{\includegraphics[width=0.16\linewidth,trim=13.cm 15.cm 15.cm 0.cm, clip]{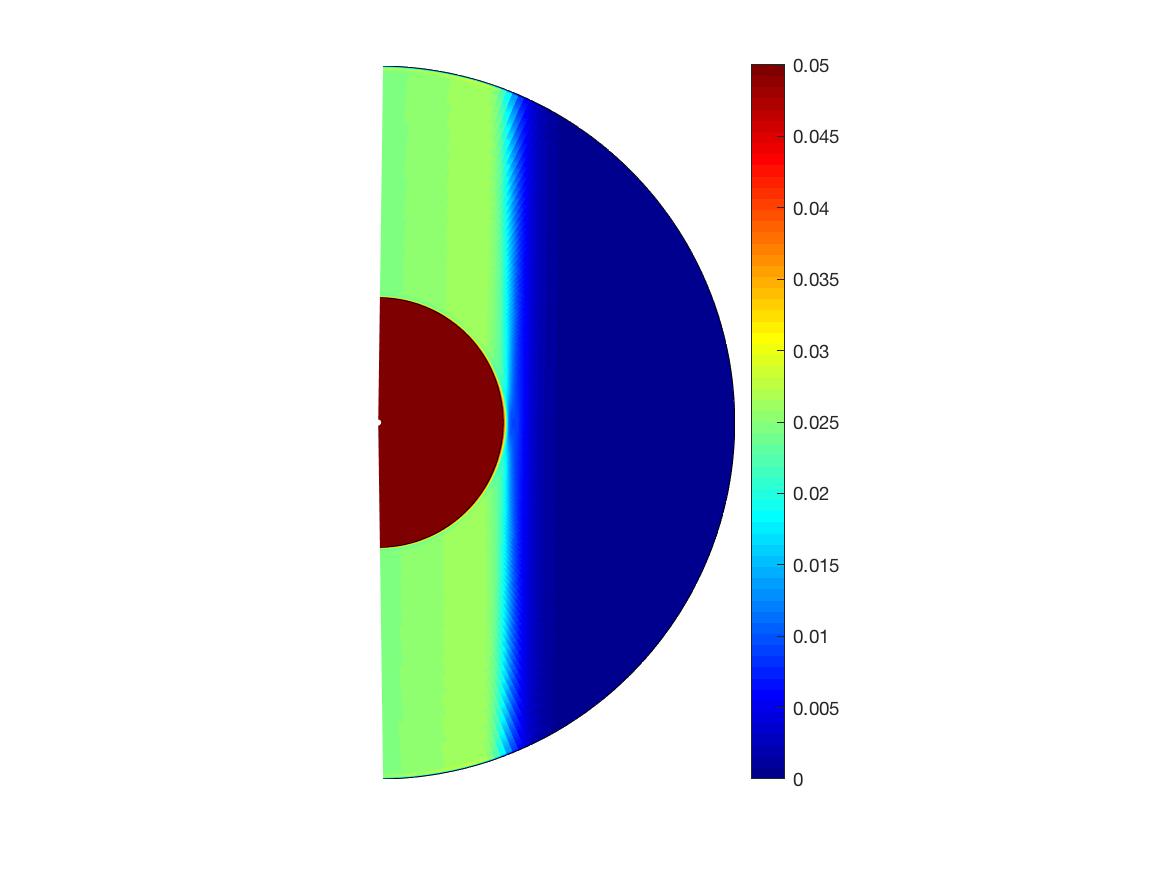}} 
   & 
\subfigure[$Q=0.15$]{\includegraphics[width=0.16\linewidth,trim=13.cm 15.cm 15.cm 0.cm, clip]{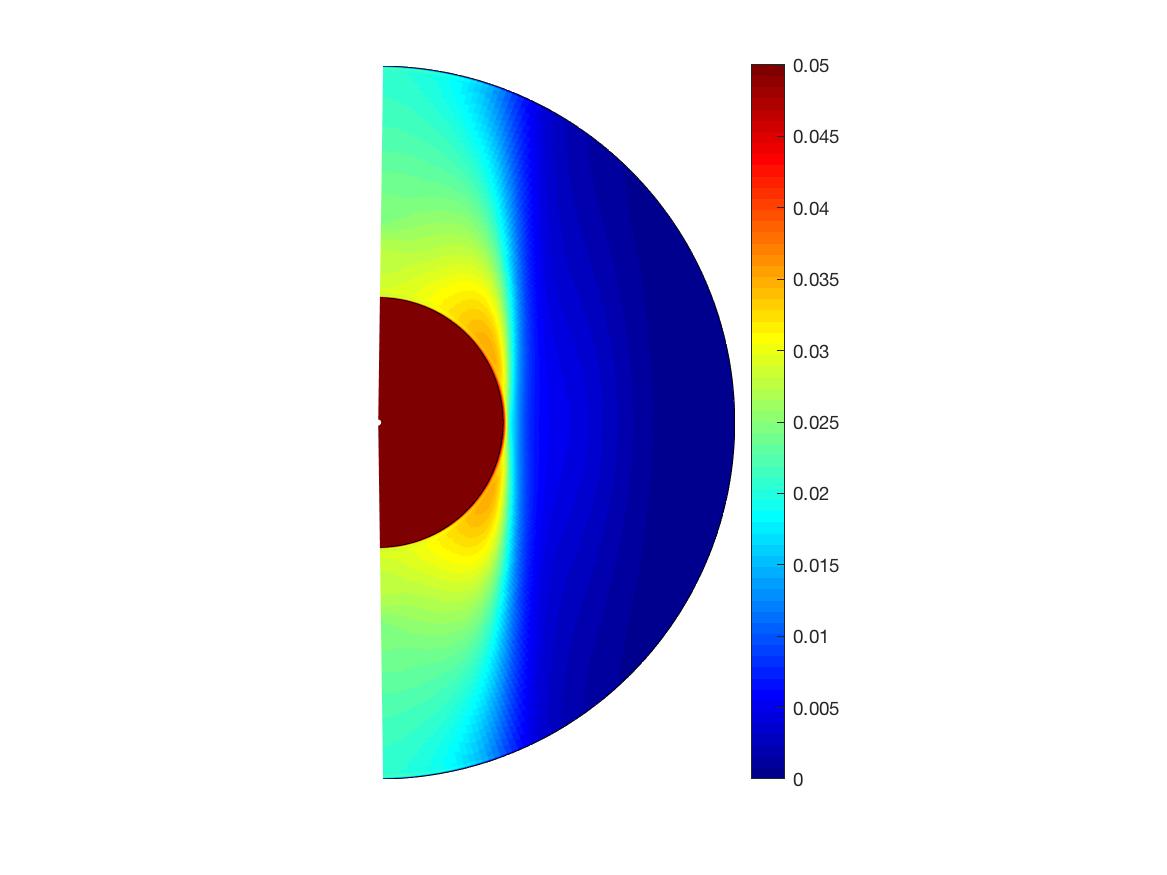}} 
   &
\subfigure[$Q=0.3$]{\includegraphics[width=0.16\linewidth,trim=13.cm 15.cm 15.cm 0.cm, clip]{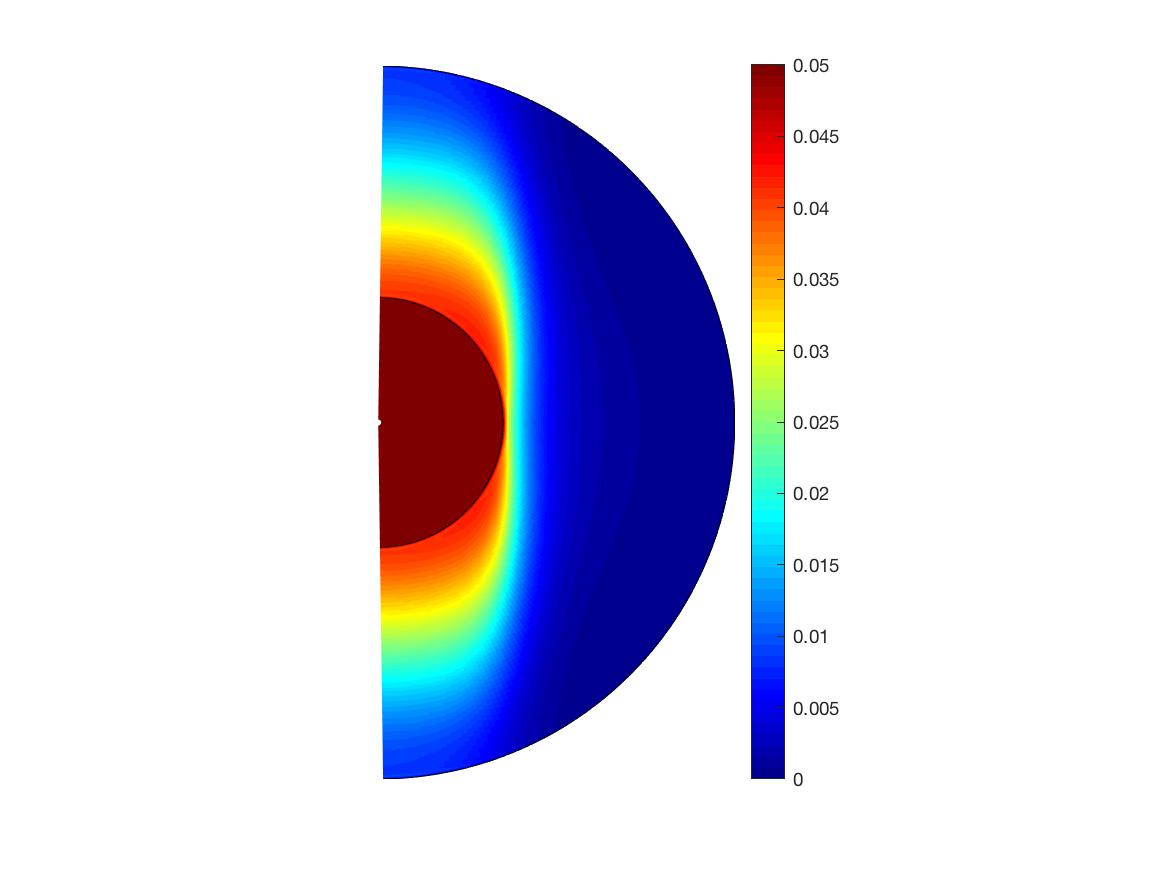}} &
\subfigure[$Q=1.5$]{\includegraphics[width=0.16\linewidth,trim=13.cm 15.cm 15.cm 0.cm, clip]{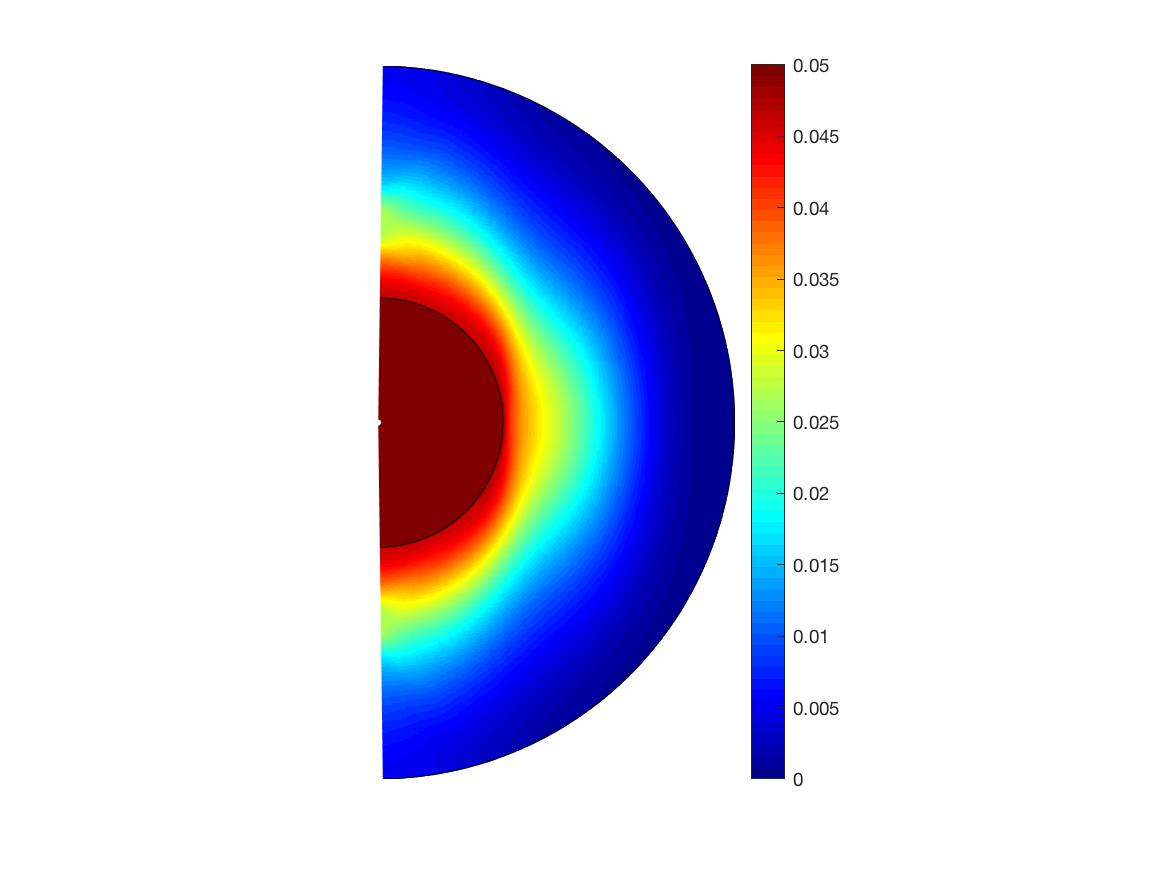}}
 &
\subfigure[$Q=3$]{\includegraphics[width=0.16\linewidth,trim=13.cm 15.cm 15.cm 0.cm, clip]{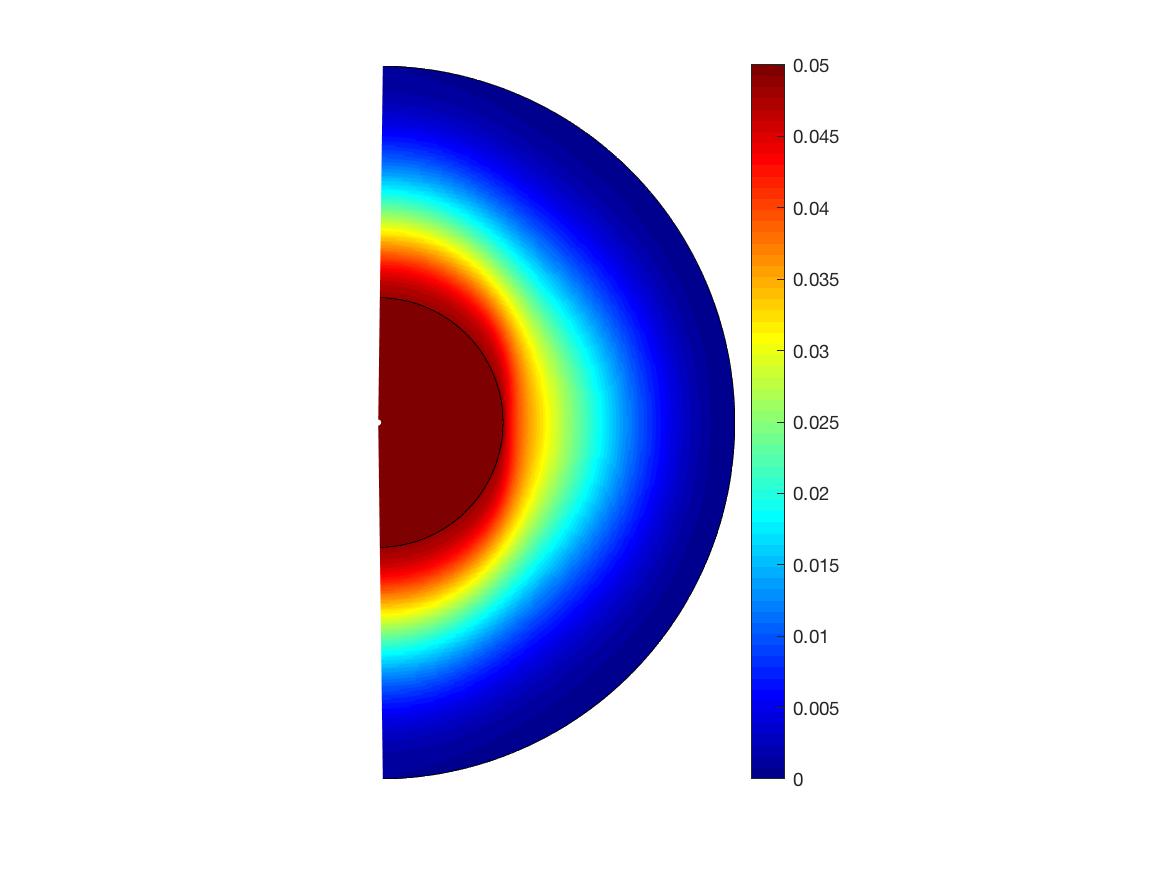}} 
&  \multirow{3}{*}{\includegraphics[width=0.08\linewidth]{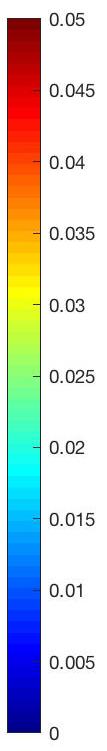}} \\
\rotatebox{90}{{\footnotesize $E=10^{-6}$, $Pr=0.05$}} &
  \subfigure[$Q=8\cdot 10^{-5}$]{\includegraphics[width=0.16\linewidth,trim=13.cm 15.cm 15.cm 0.cm, clip]{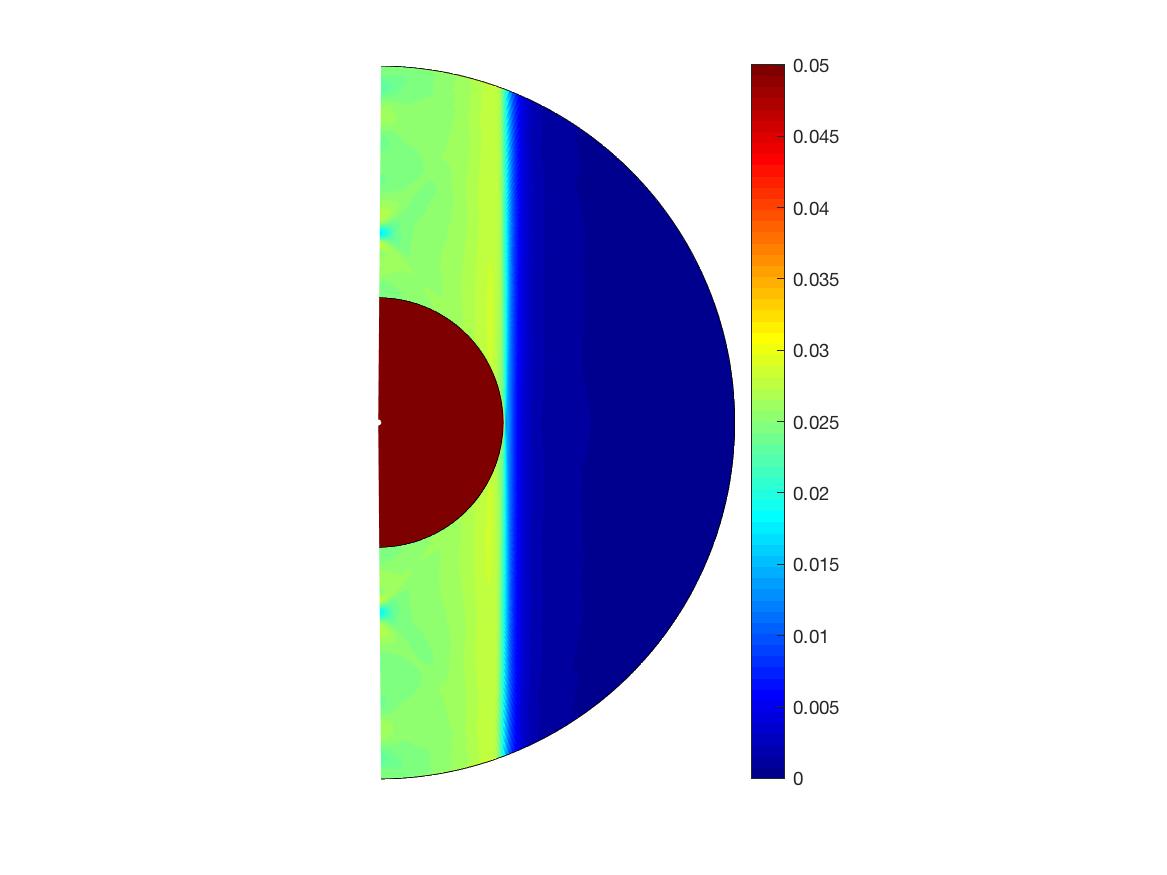}} 
 &
\subfigure[$Q=0.015$]{\includegraphics[width=0.16\linewidth,trim=13.cm 15.cm 15.cm 0.cm, clip]{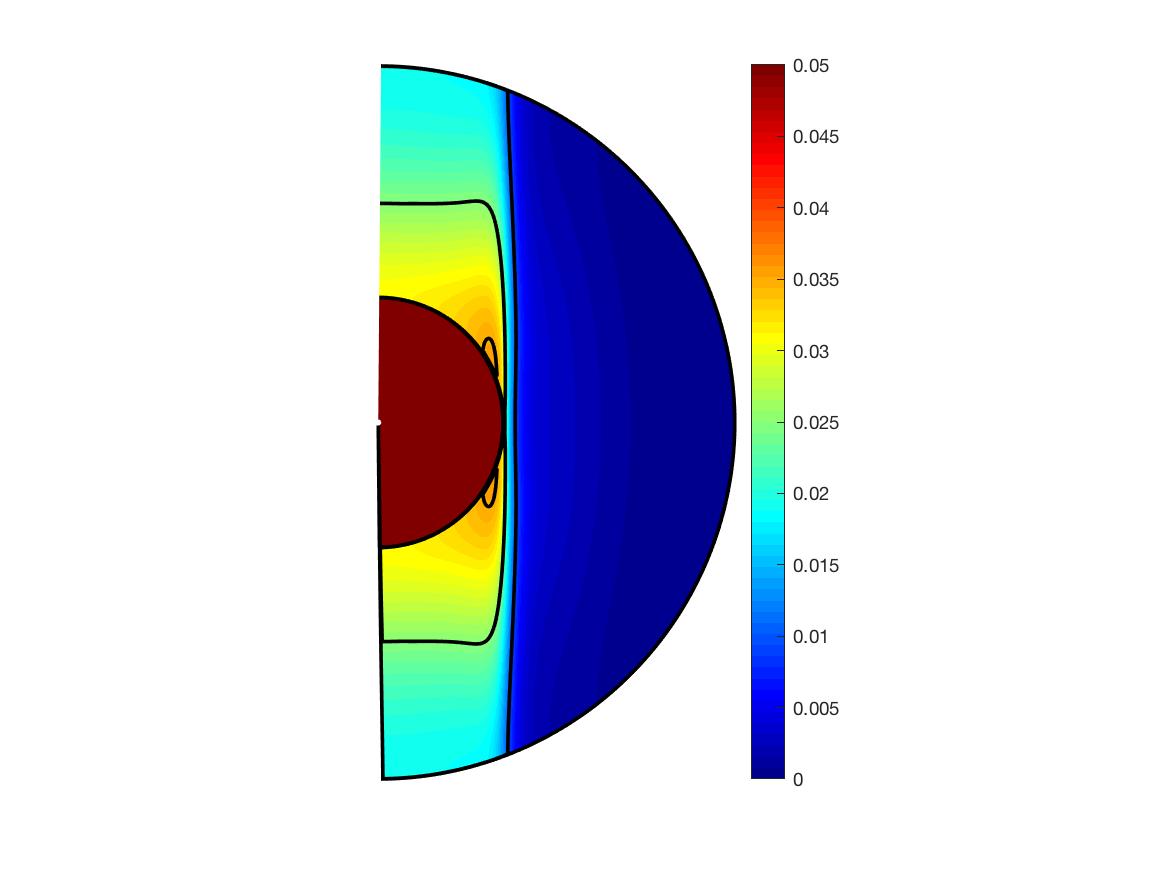}}
 &
\subfigure[$Q=0.15$]{\includegraphics[width=0.16\linewidth,trim=13.cm 15.cm 15.cm 0.cm, clip]{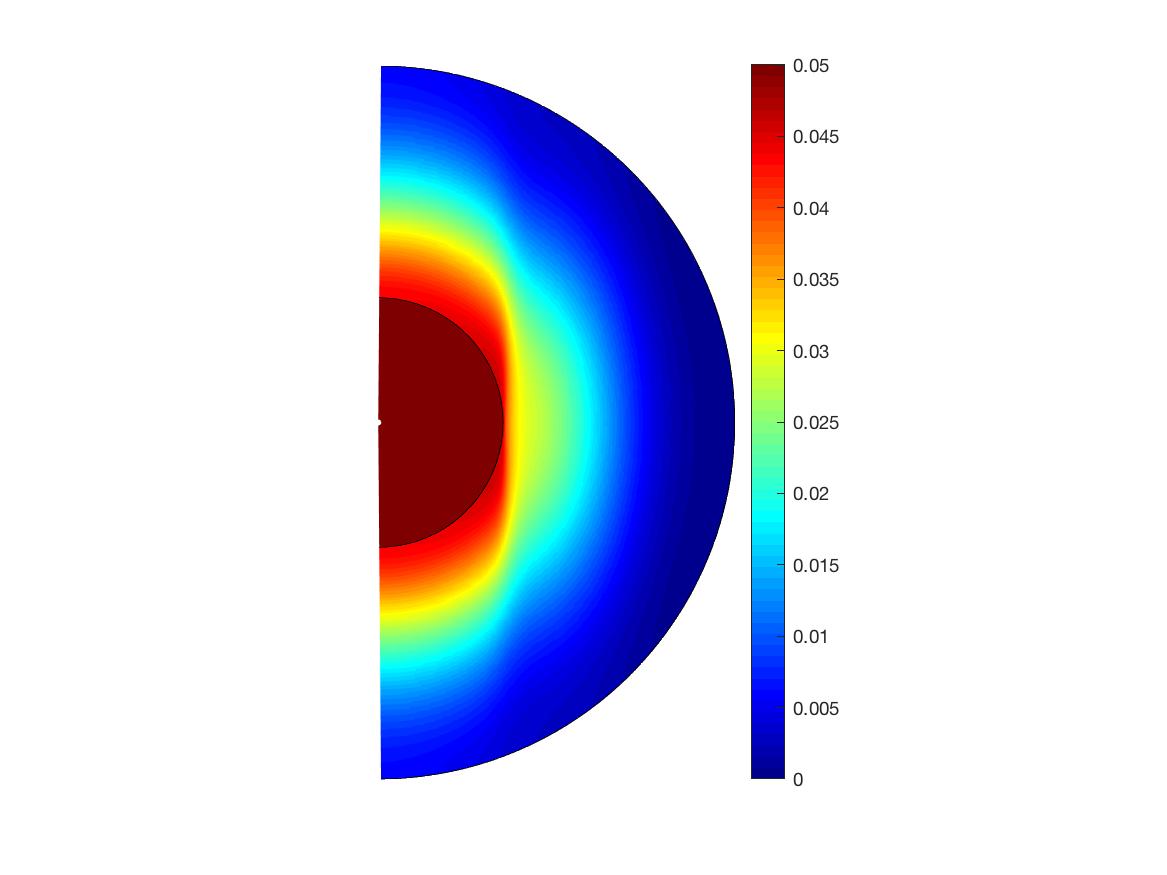}} &
\subfigure[$Q=1.5$]{\includegraphics[width=0.16\linewidth,trim=13.cm 15.cm 15.cm 0.cm, clip]{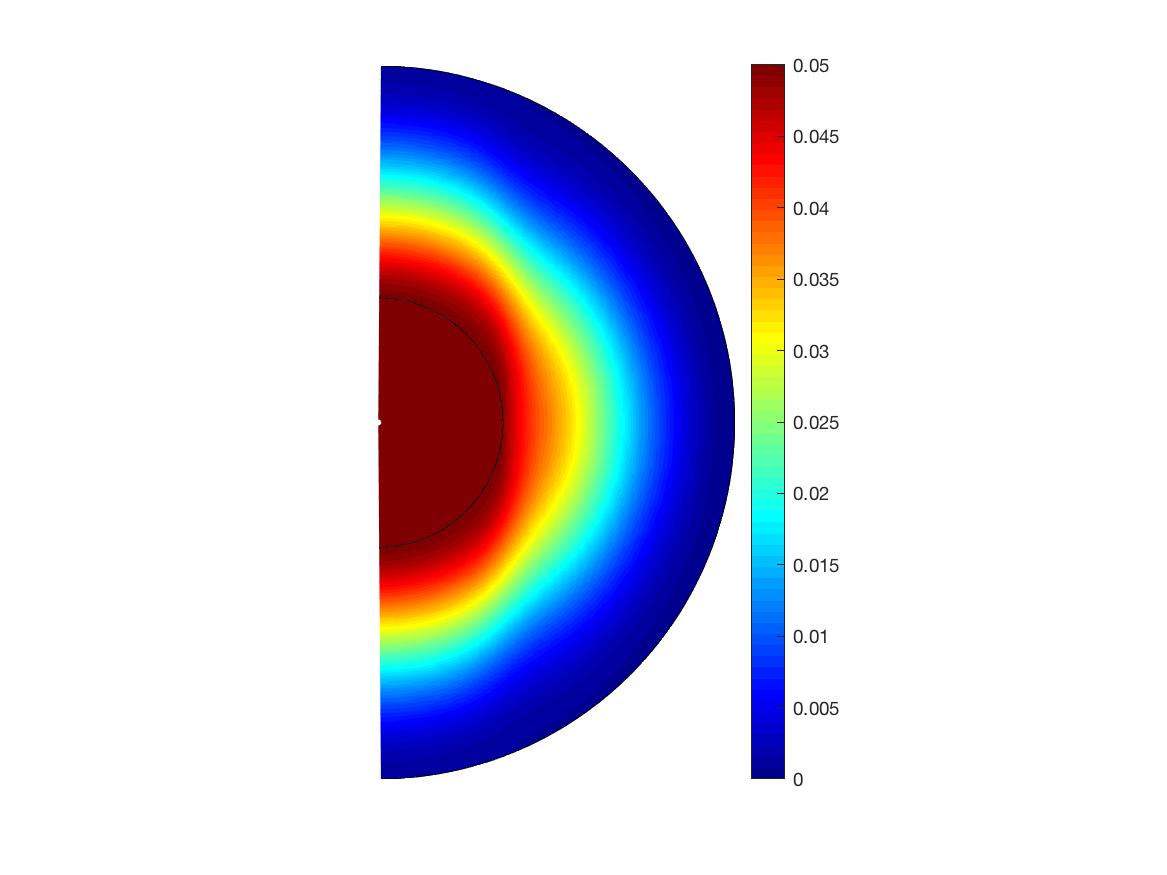}}&
\subfigure[$Q=3$]{\includegraphics[width=0.16\linewidth,trim=13.cm 15.cm 15.cm 0.cm, clip]{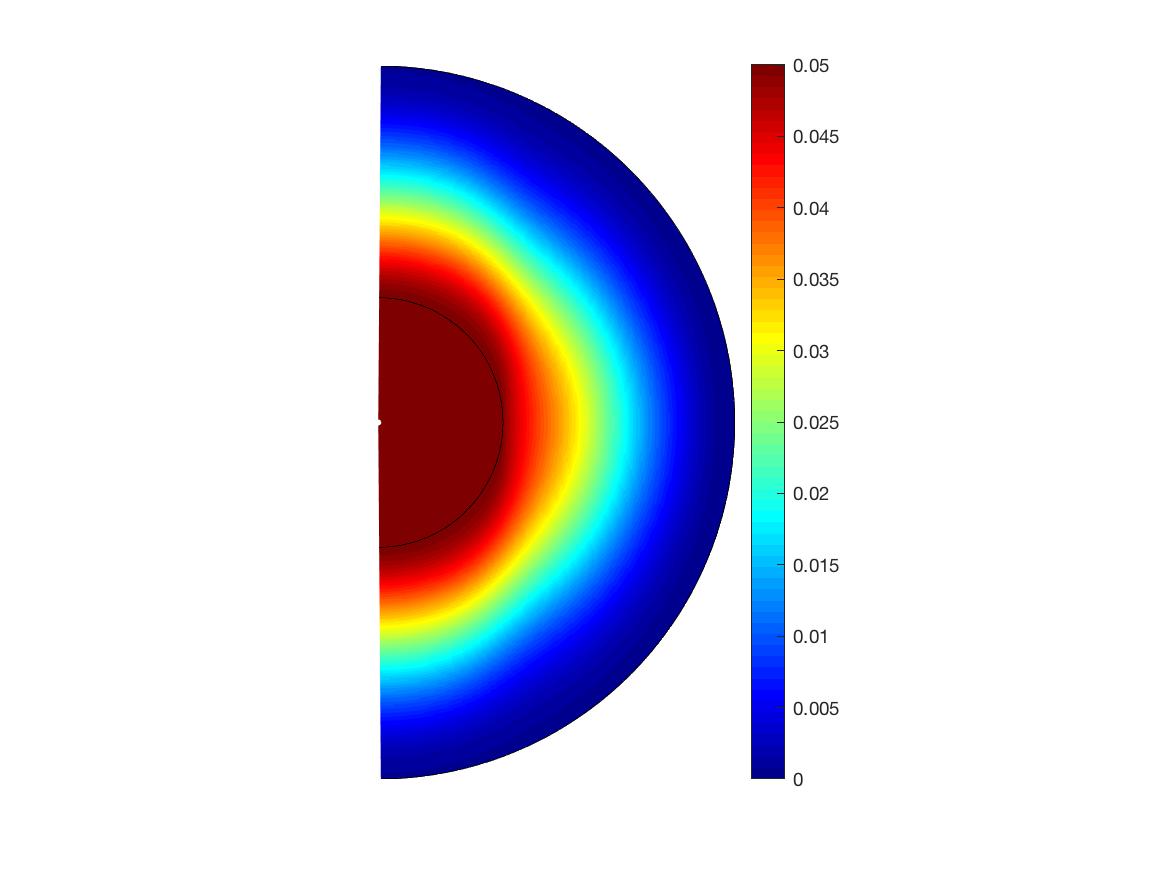}}
&  \\
\rotatebox{90}{{\footnotesize $E=10^{-5}$, $Pr=0.05$}} &
  \subfigure[$Q=5\cdot 10^{-4}$]{\includegraphics[width=0.16\linewidth,trim=13.cm 15.cm 15.cm 0.cm, clip]{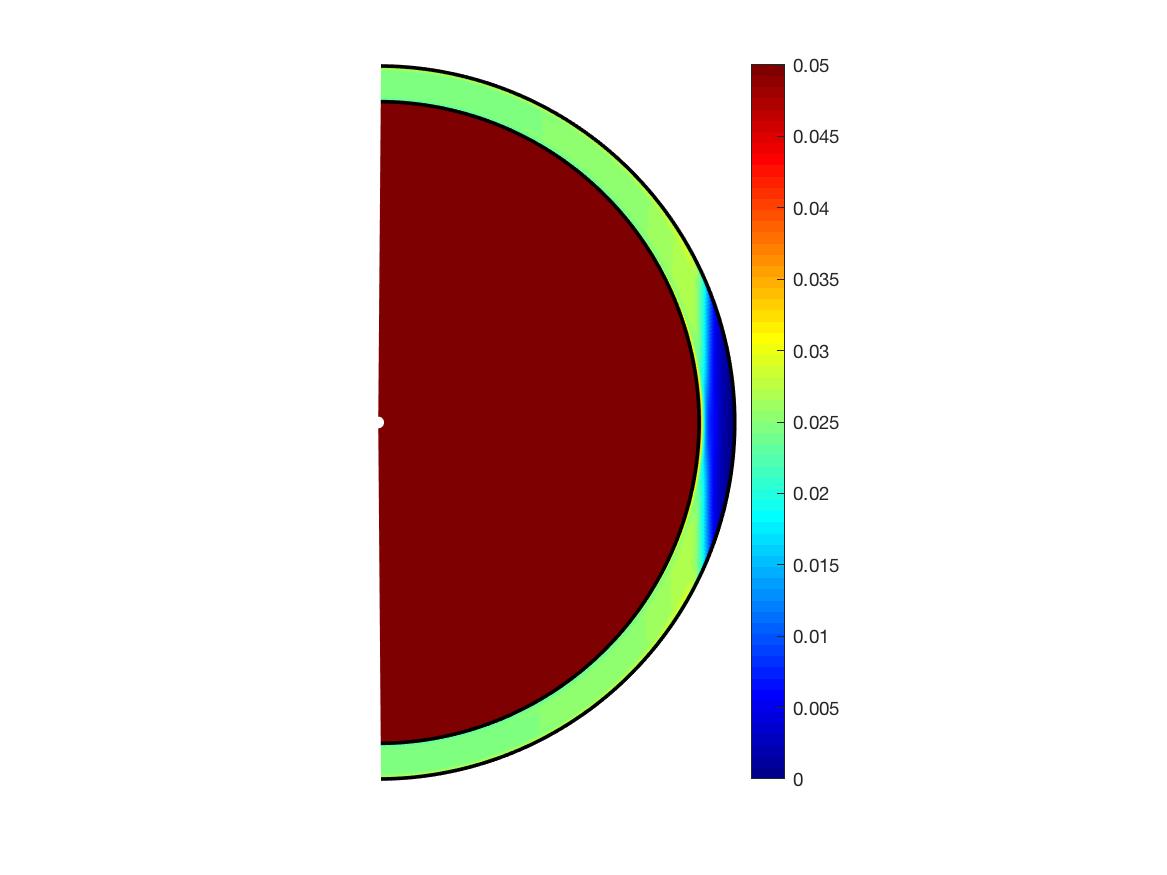}} 
 &
\subfigure[$Q=0.1$]{\includegraphics[width=0.16\linewidth,trim=13.cm 15.cm 15.cm 0.cm, clip]{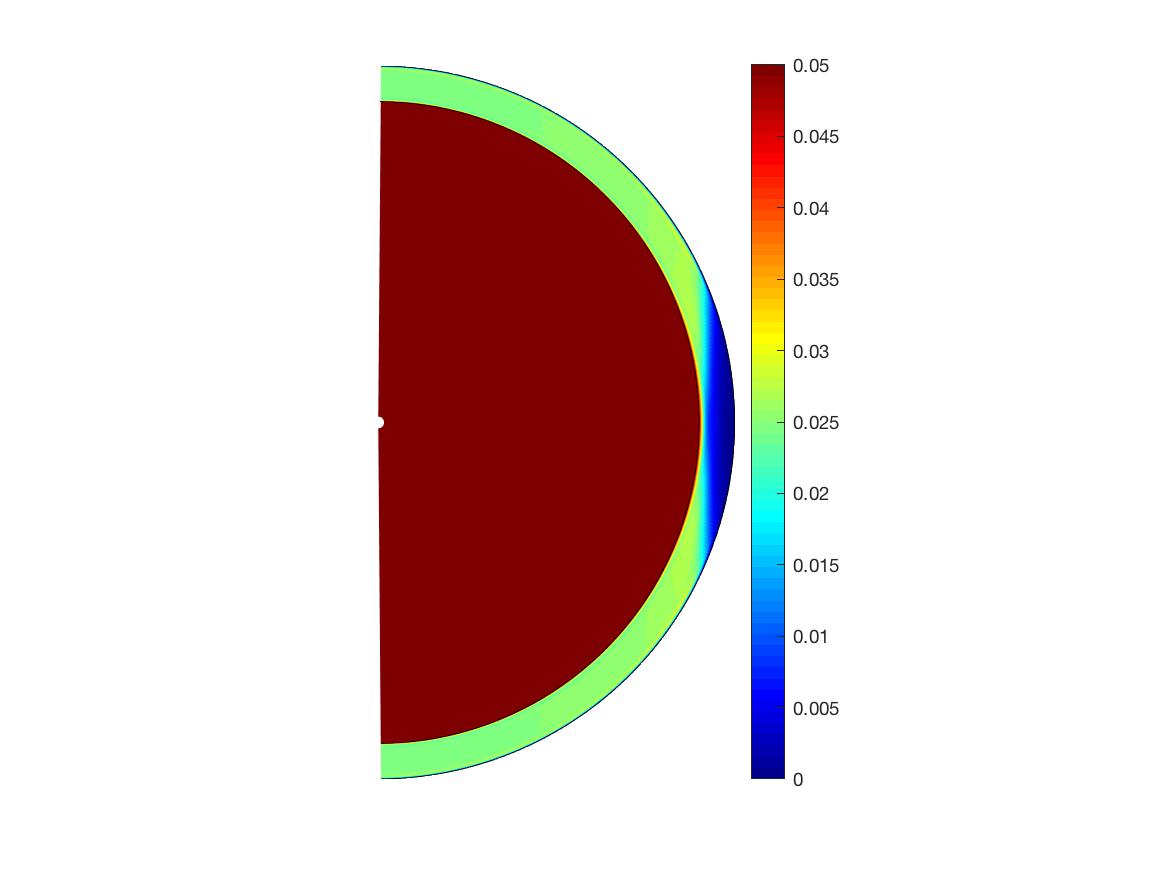}}
 &
\subfigure[$Q=1$]{\includegraphics[width=0.16\linewidth,trim=13.cm 15.cm 15.cm 0.cm, clip]{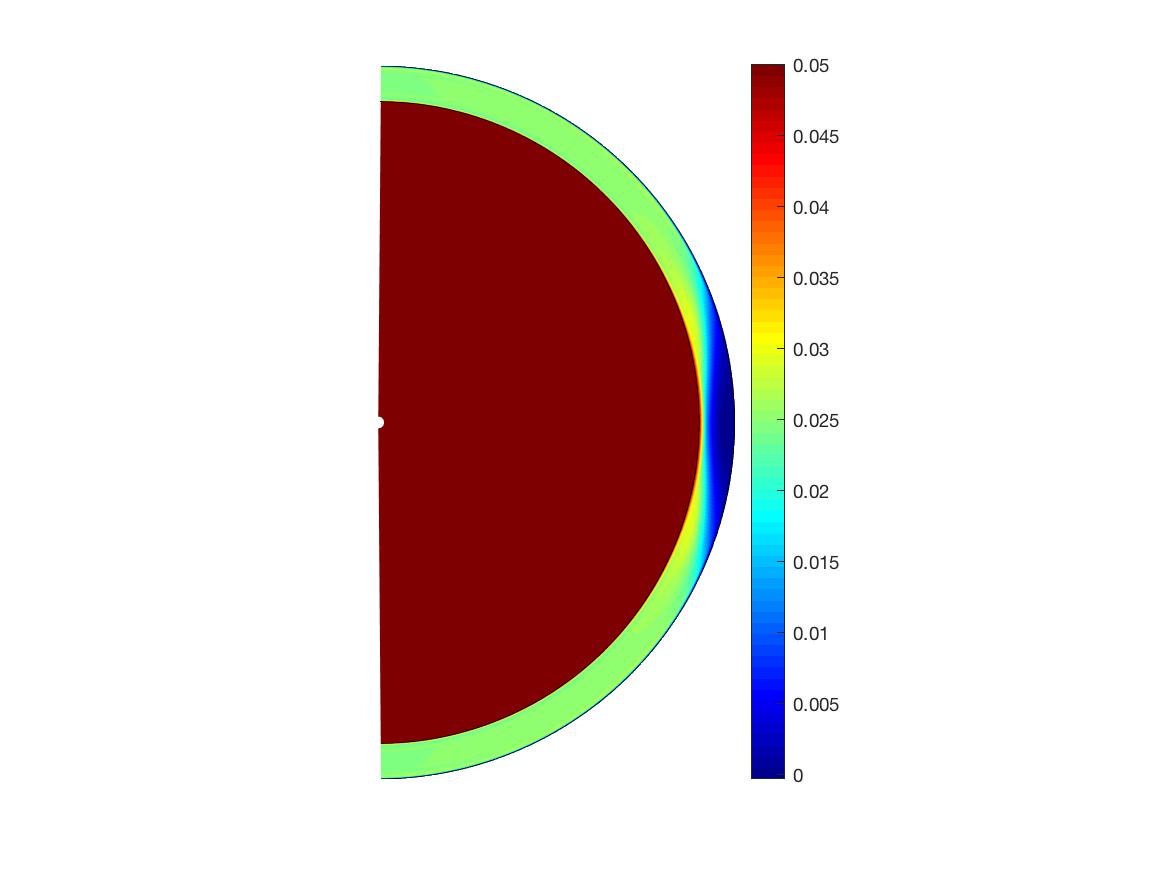}} &
\subfigure[$Q=10$]{\includegraphics[width=0.16\linewidth,trim=13.cm 15.cm 15.cm 0.cm, clip]{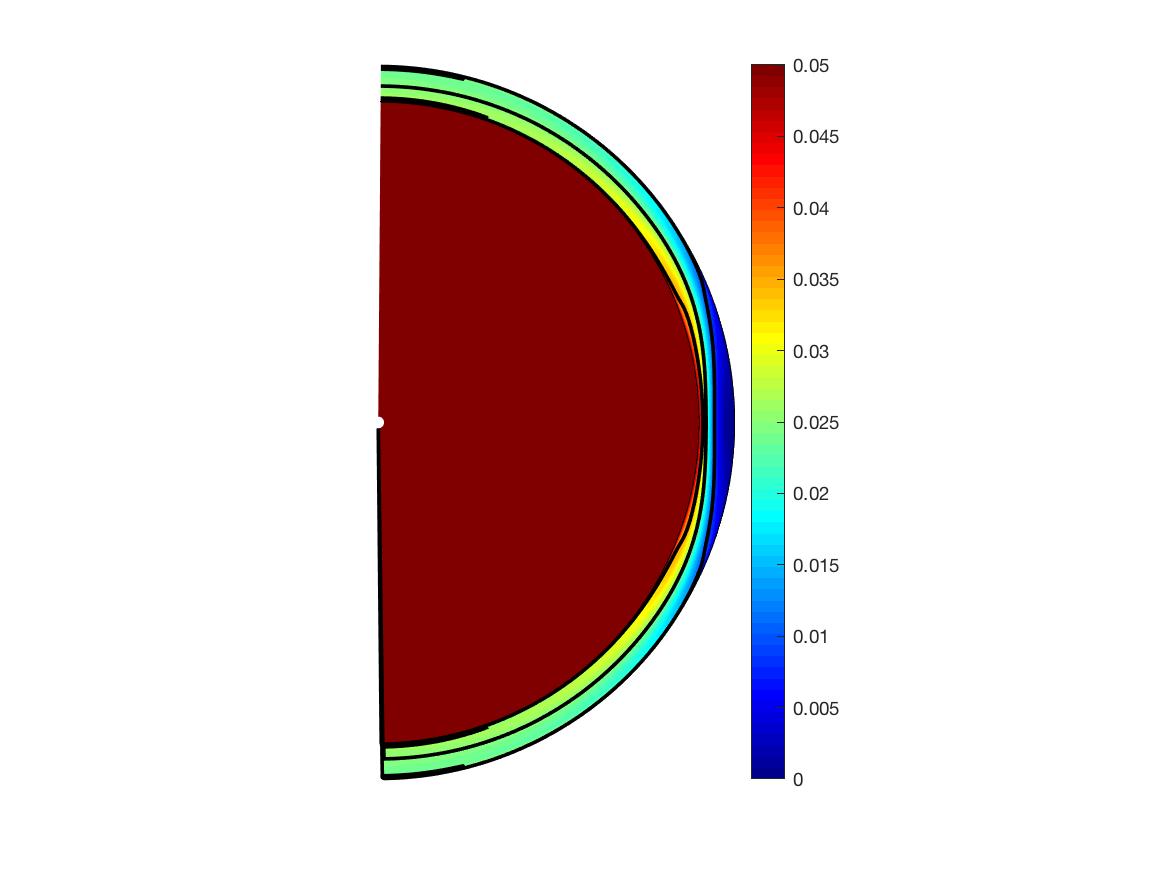}}&
\subfigure[$Q=20$]{\includegraphics[width=0.16\linewidth,trim=13.cm 15.cm 15.cm 0.cm, clip]{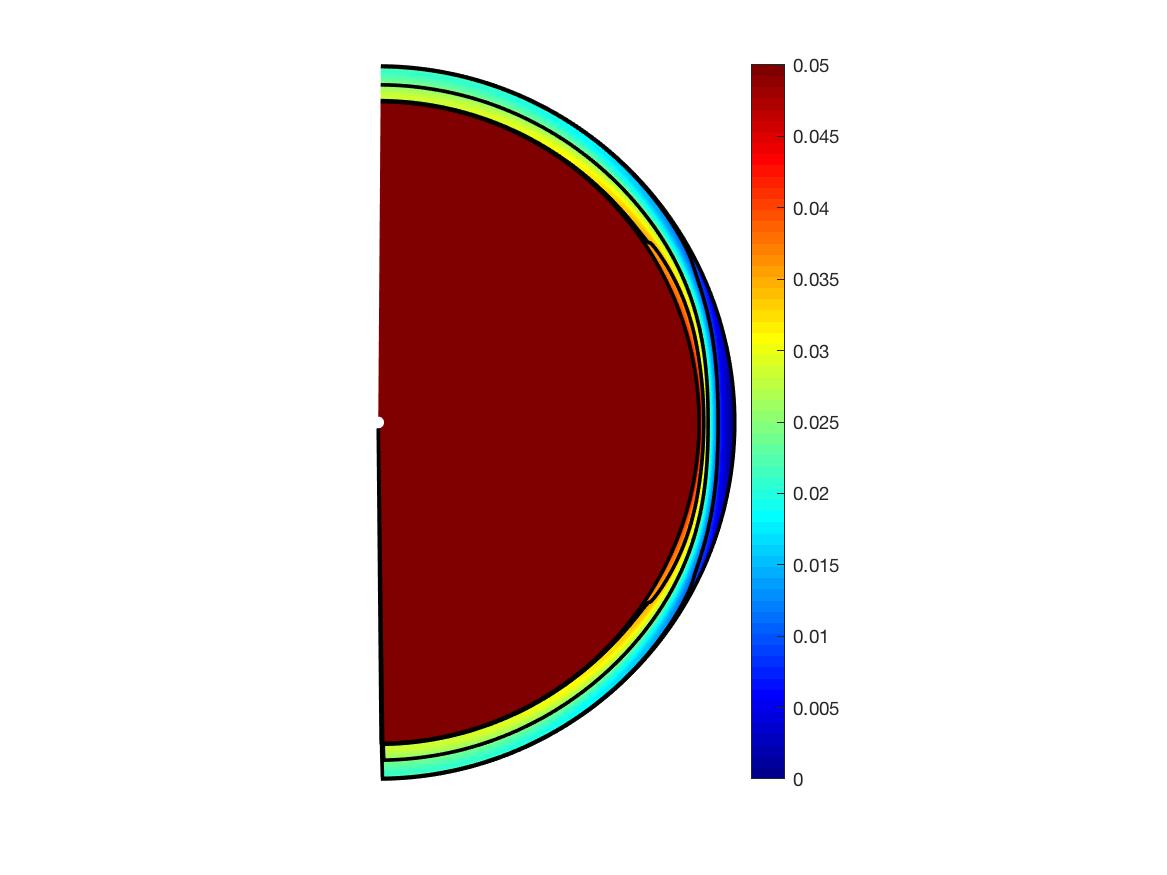}}
 & \\
\rotatebox{90}{{\footnotesize $E=10^{-6}$, $Pr=0.05$}} &
  \subfigure[$Q=10^{-4}$]{\includegraphics[width=0.16\linewidth,trim=13.cm 15.cm 15.cm 0.cm, clip]{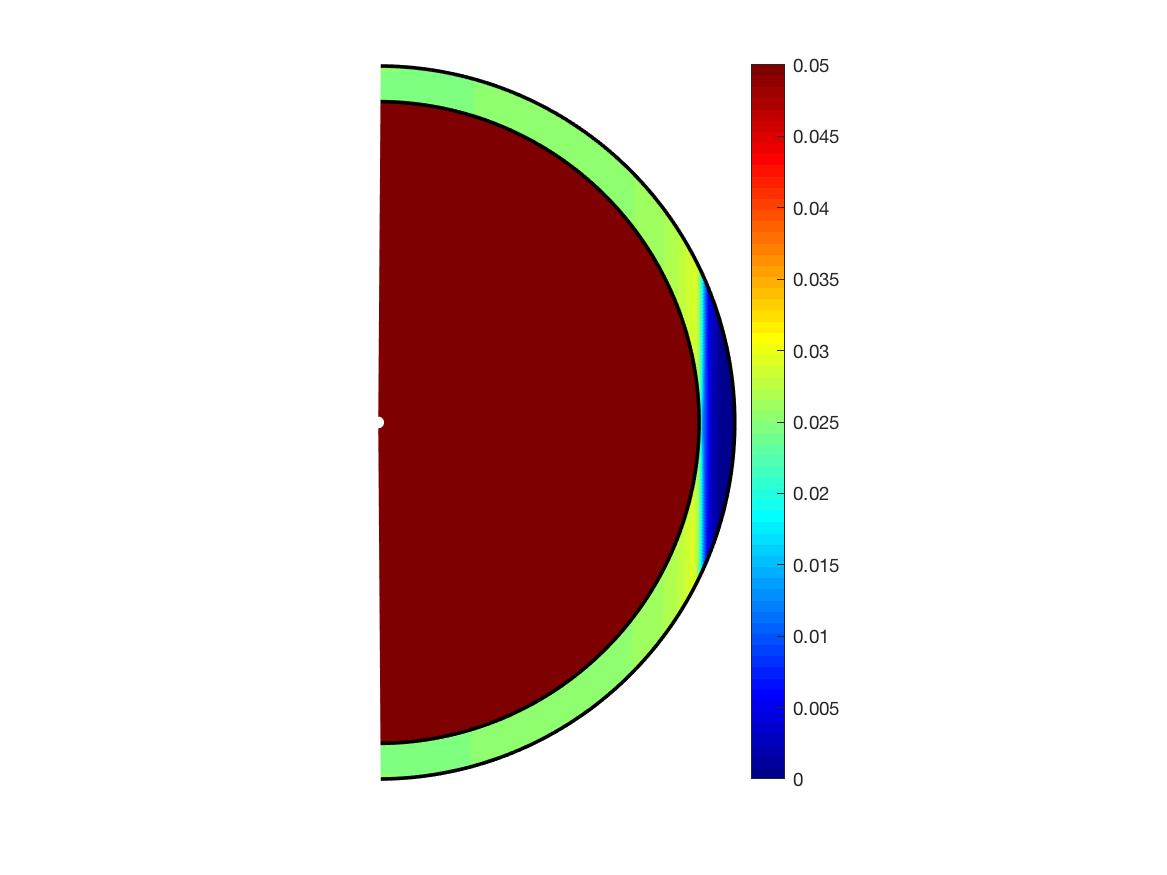}} 
 &
\subfigure[$Q=0.1$]{\includegraphics[width=0.16\linewidth,trim=13.cm 15.cm 15.cm 0.cm, clip]{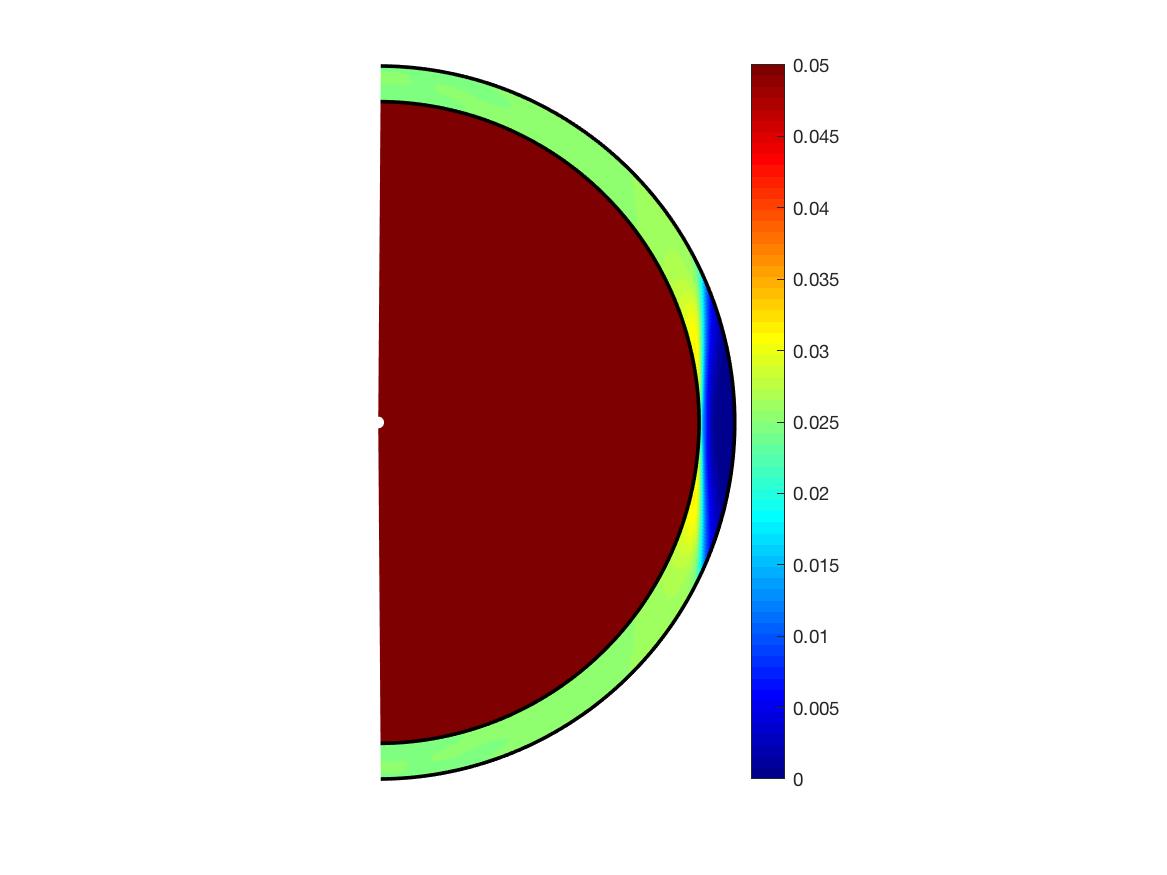}}
 &
\subfigure[$Q=1$]{\includegraphics[width=0.16\linewidth,trim=13.cm 15.cm 15.cm 0.cm, clip]{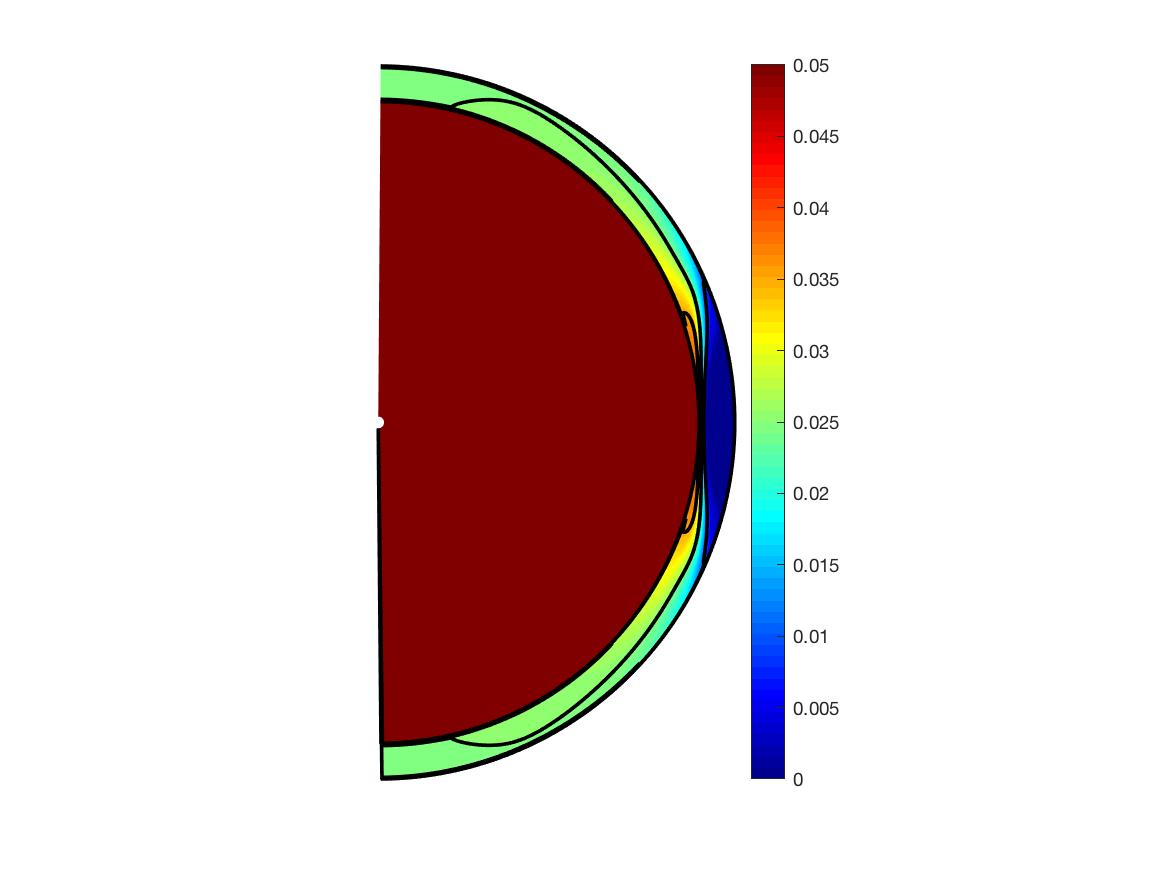}} &
\subfigure[$Q=10$]{\includegraphics[width=0.16\linewidth,trim=13.cm 15.cm 15.cm 0.cm, clip]{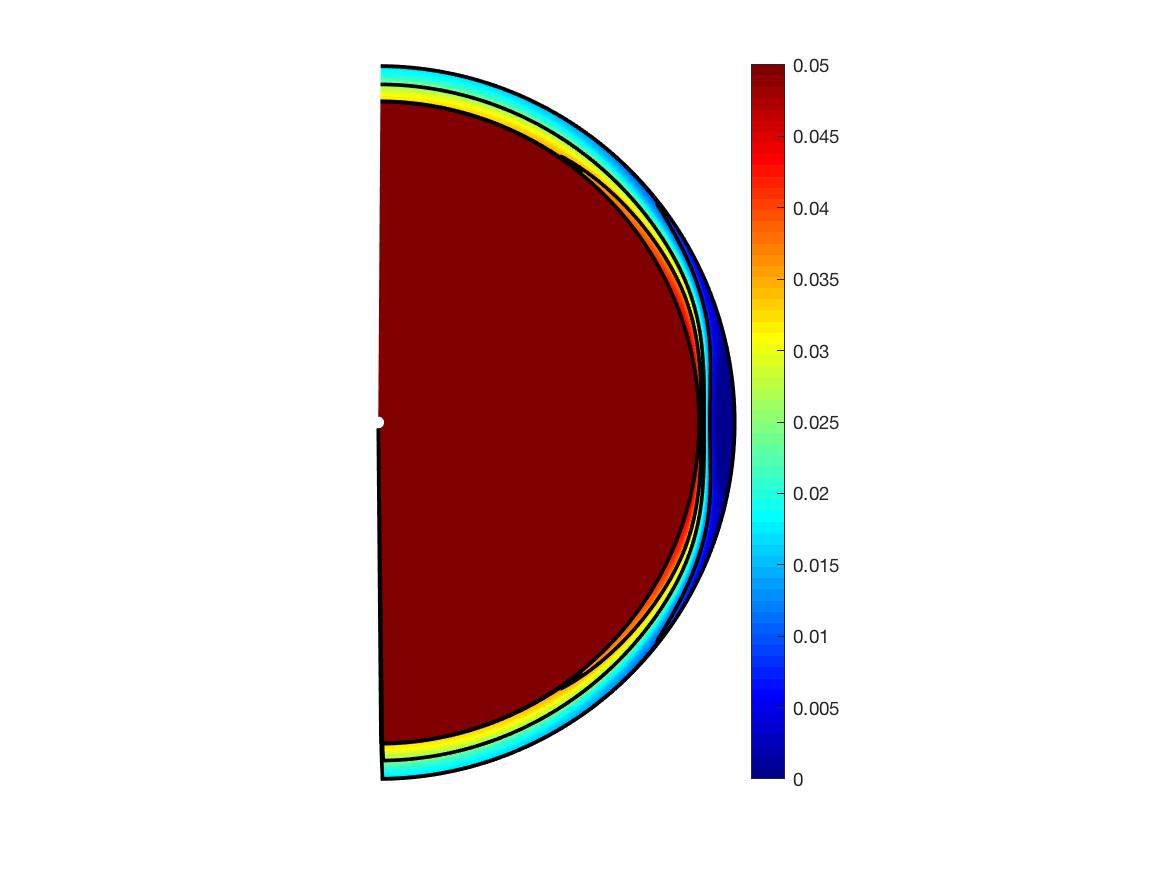}}&
\subfigure[$Q=20$]{\includegraphics[width=0.16\linewidth,trim=13.cm 15.cm 15.cm 0.cm, clip]{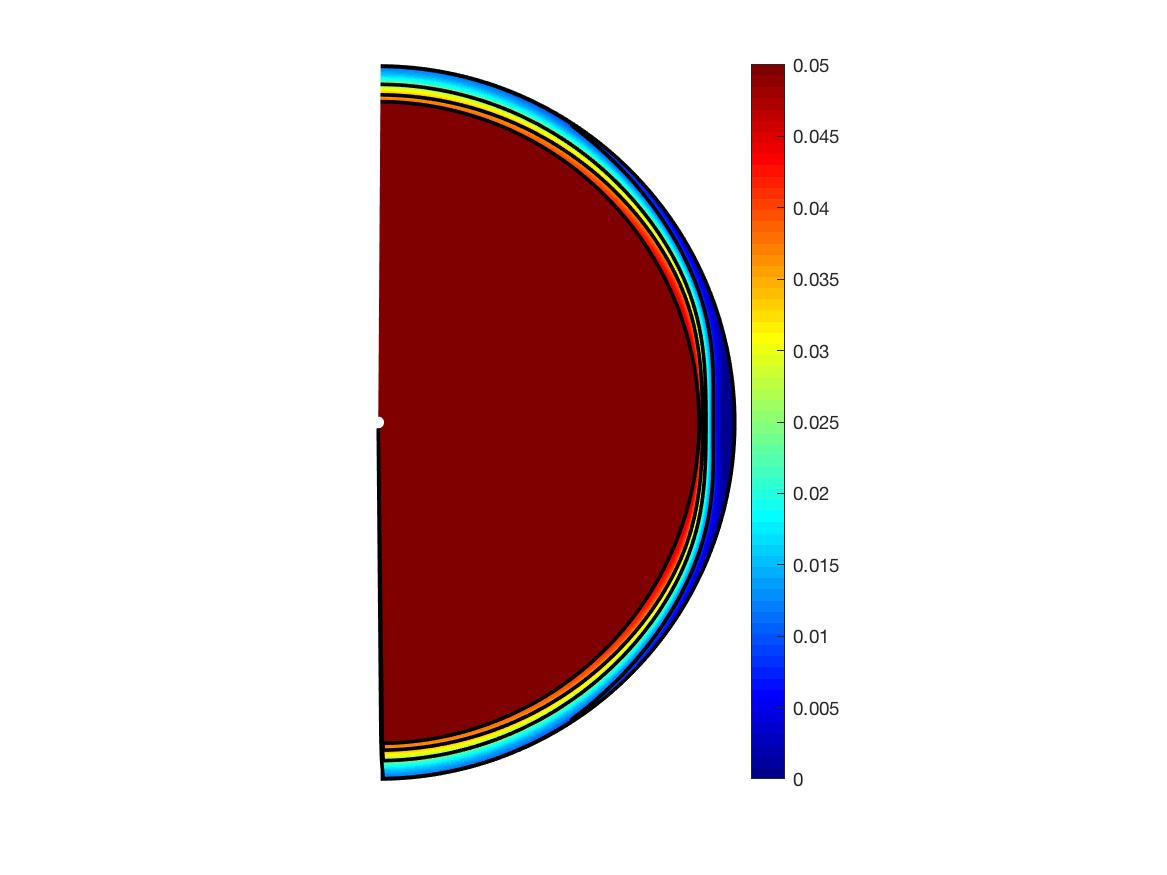}}
 & \\
\multicolumn{6}{c}{$\overrightarrow{\;\;\;\;\;\;\;\;\;\;\; Pr\left(\frac{N}{\Omega}\right)^2\equiv Pr\,E\tilde{R}a\left(\dfrac{r_o}{H}\right)\equiv Q\;\;\;\;\;\;\;\;\;}$}
\end{tabular}
\captionof{figure}{Rotation profile $\Omega(r, \theta)$ in the frame of reference of the outer sphere for purely hydrodynamic Couette flows with a weak differential rotation ($Ro = 0.05$) for different values of $E$ and $Pr$, two different aspect ratios ($r_i/r_o = 0.35$ for the first two rows, $0.9$ for the last two), as well as increasing values of $Q$ (from left to right). As the amplitude of stratification is increased, the flow undergoes a continuous transition from cylindrical to spherical symmetry. Earth's stratified layer lies in the intermediate regime, while all regimes may be relevant for stellar radiative zones. The solid black lines shown in non-trivial geometries are contours of equal rotation rate.}
\label{fig:hydro_flows}
\end{figure*}

In order to carry out the study of this system, we follow what has been done in previous studies \citep{dormy98, dormy02} and consider a weak differential rotation, which corresponds to negligible non-linear terms. Thus, we have linearised equation \eqref{eq:full_set} by expressing the magnetic field as $\bm{B} = \bm{B_0} + \bm{b}$ ($\bm{B_0}$ being a dipolar magnetic field, whose amplitude is controlled by $\Lambda$), the temperature as $T = T_s + \Theta$ (where $T_s$ is the stationary solution to the heat equation without source $\Delta T_s = 0$), and by neglecting the advection term in the Navier-Stokes equation. The system of equations
stands thus:
\bse
\begin{align}
\dfrac{\upartial\bm{v}}{\upartial t} \,=\,&\, E\Delta \bm{v} - 2\bm{e_z}\times\bm{v} - \bm{\nabla} P \nonumber\\
&\, + \dfrac{E\Lambda}{Pm}((\bm{\nabla}\times\bm{B_0})\times\bm{b} + (\bm{\nabla}\times\bm{b})\times\bm{B_0}) + E\widetilde{Ra}\Theta\bm{e_r}\,,\\
\dfrac{\upartial\bm{b}}{\upartial t}\, =\, &\,\dfrac{E}{Pm}\Delta\bm{b} + {\bm\nabla}\times(\bm{v}\times\bm{B_0})\,, \\
\dfrac{\upartial \Theta}{\upartial t}\, =\, &\,-\,(\bm{v}{\bm\cdot}\bm{\nabla}) T_s + \dfrac{E}{Pr}\Delta \Theta \,,
\end{align}
\ese
supported by the boundary conditions
\begin{equation}
\begin{array}{l}
\bm{v}(r_i) = \Omega_i r_i \bm{e_\phi} \\
\bm{v}(r_o) = \bm{0} \\
\Theta(r_i) = \Theta(r_o) = 0~,
\end{array}
\end{equation}
while 
%
%
the boundary conditions on the magnetic field $\bm{b}$ are obtained by solving the induction equation within the inner boundary, and by considering that $\bm{b}$ is a potential field outside the outer boundary.

We have then integrated these linearised equations using the PaRoDy code \citep[][and later collaborations]{dormy98}. We checked the temporal stationarity of our steady solutions manually, by waiting for the output physical quantities (fluid velocity, magnetic field, temperature) to stop varying with time. On the other hand, we ensured that spatial resolution was sufficient by checking that the obtained steady states were not subject to change under a significant increase of spatial resolution; furthermore, we ensured that the energy of the modes of highest angular degree was at least three orders of magnitude lower than for the lowest angular degree. Note that compared to previous studies \citep{dormy98,dormy02}, the novelty here lies in the introduction of stable stratification. Otherwise, we have followed their numerical method.

\section{Hydrodynamical results and symmetry of the flow}\label{sec:hydro}

We will discuss in detail the full MHD results in the next section; however, it is instructive to first describe the effect of a stable stratification on a spherical Couette flow in the purely hydrodynamical regime, that is with $\Lambda = 0$. Previous studies \citep{barcilon67a,barcilon67b} investigated such an effect in a cylindrical geometry, and found that although the Rayleigh number and the Ekman number are independent control parameters, the structure of the flow is solely controlled by the parameter $Q \equiv Pr\times (N / \Omega_o)^2$ where $N$ is the Br\"unt-V\"ais\"al\"a frequency, which we estimate by $N^2=\alpha g\Delta T/H$ (where $H = r_o - r_i$). Using the dimensionless parameters introduced above, we can write $Q=Pr\times E\times \widetilde{R}a\times (r_o / H)$. Note that the control parameter $Q$ is tightly related to the \textit{convective} Rossby number defined as the ratio of the rotation period to the buoyancy free-fall time $(N / \Omega_o)$. \citet{garaud02} has studied numerically a non-linear version of this problem (including centrifugal effect and density stratification) and noticed the important role of $Q$. This was also the case in the initial value version of this problem studied numerically by \citet{gauratPHD}. We found a similar dependence  with $Q$ in our data. More precisely, $3$ different regimes can be observed depending on the value of $Q$, as illustrated by figure \ref{fig:hydro_flows} \citep{barcilon67a,barcilon67b}. If $Q \ll E^{2/3}$ (left panels in figure \ref{fig:hydro_flows}), the flow is driven by Coriolis forces and exhibits a cylindrical geometry. On the other hand, if $Q \gg 1$ (right panels in figure \ref{fig:hydro_flows}), the flow is driven by buoyancy, and a spherical symmetry is observed for the flow. In between, neither one dominates the other. To draw figure \ref{fig:hydro_flows}, we have varied the value of $Q$ by changing $E$, $Pr$ and the aspect ratio. It is important to note that additional simulations (not shown) have been run with different parameters and also show that $Q$ is the key parameter.

Due to the smallness of $Pr$ (less than $10^{-5}$ according to \citealt{book_dormy} and \citealt{thomasWeiss}), $Q$ would be of the order $1$ in the solar radiative zone. Observations of stellar rotation periods show that $\Omega$ can vary by several orders of magnitude \citep{donati09} and the different regimes highlighted here are all relevant in the stellar context. 

For the Earth case, \citet{robertsK13} have provided an estimate for the Prandtl number at $Pr=0.1$. However, in the outer core, the kinematic viscosity and the thermal diffusivity are so low compared to inertia that turbulence can significantly increase their effective values (turbulent diffusivities) such that $Pr$ may approach unity. \citet{buffett16} have argued on the possibility that the Brünt-Väisälä frequency $N=\sqrt{\alpha g\Delta T/H}$ may be comparable to Earth rotation rate $\Omega_o$. Consequently, the modified Rayleigh number would be linked to the Ekman number by the relationship $\widetilde{Ra}\, E \sim (H/r_o)$. In this case, the Earth would lie in the intermediate regime ($Q_\text{Earth} \sim 1$), meaning that the rotation profile is unlikely to present a simple geometry.

\begin{figure*}[t!]
\centering
\begin{tabular}{ccc}
\subfigure[$E=10^{-5}$, $\Lambda=1$]{\includegraphics[scale=0.32]{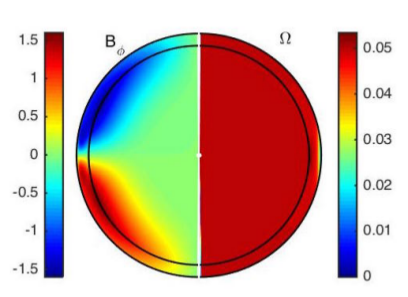}}  &
\subfigure[$E=10^{-6}$, $\widetilde{Ra}=2\cdot 10^5$]{\includegraphics[scale=0.27]{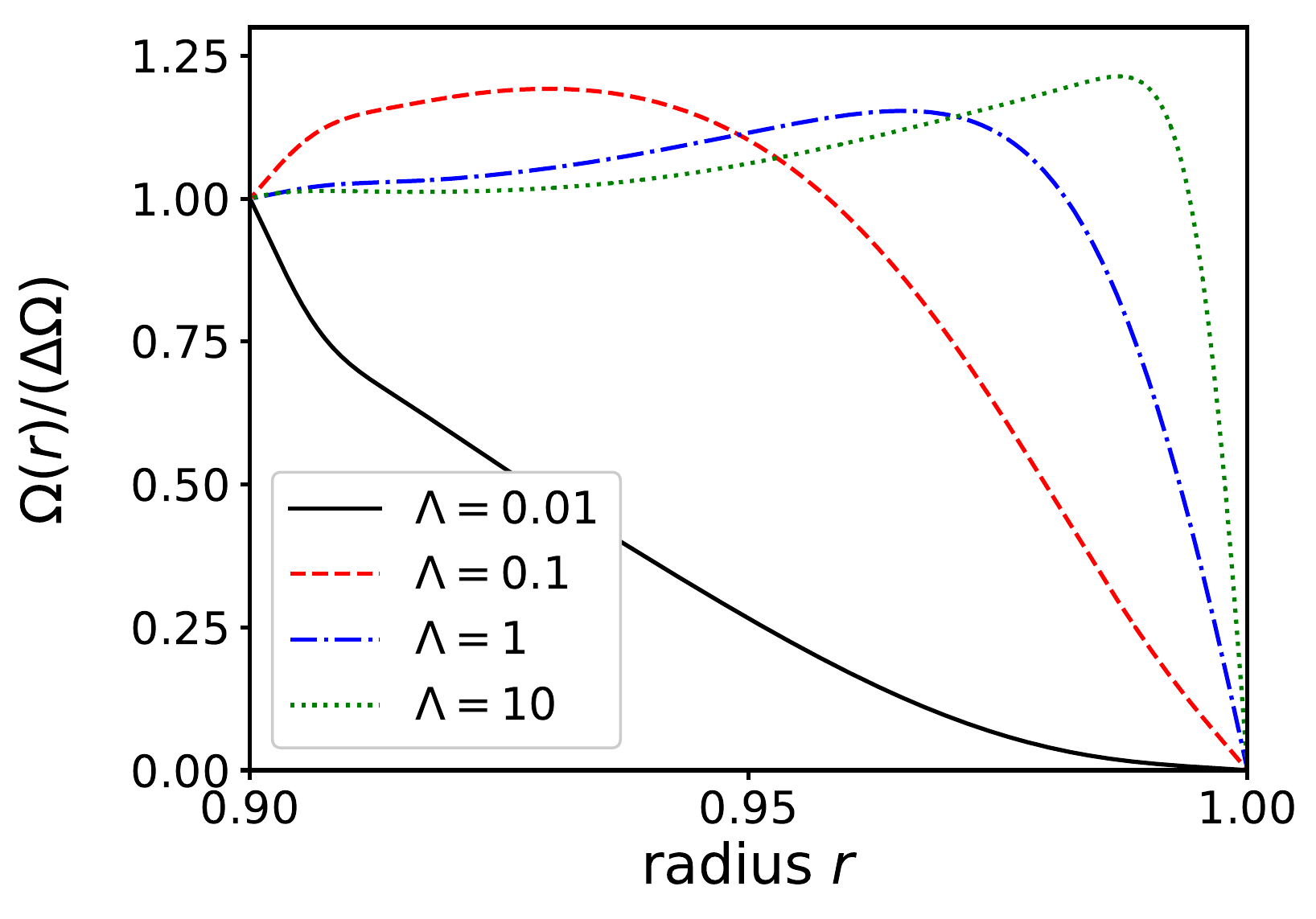}} &
\subfigure[$E=10^{-7}$]{\includegraphics[scale=0.27]{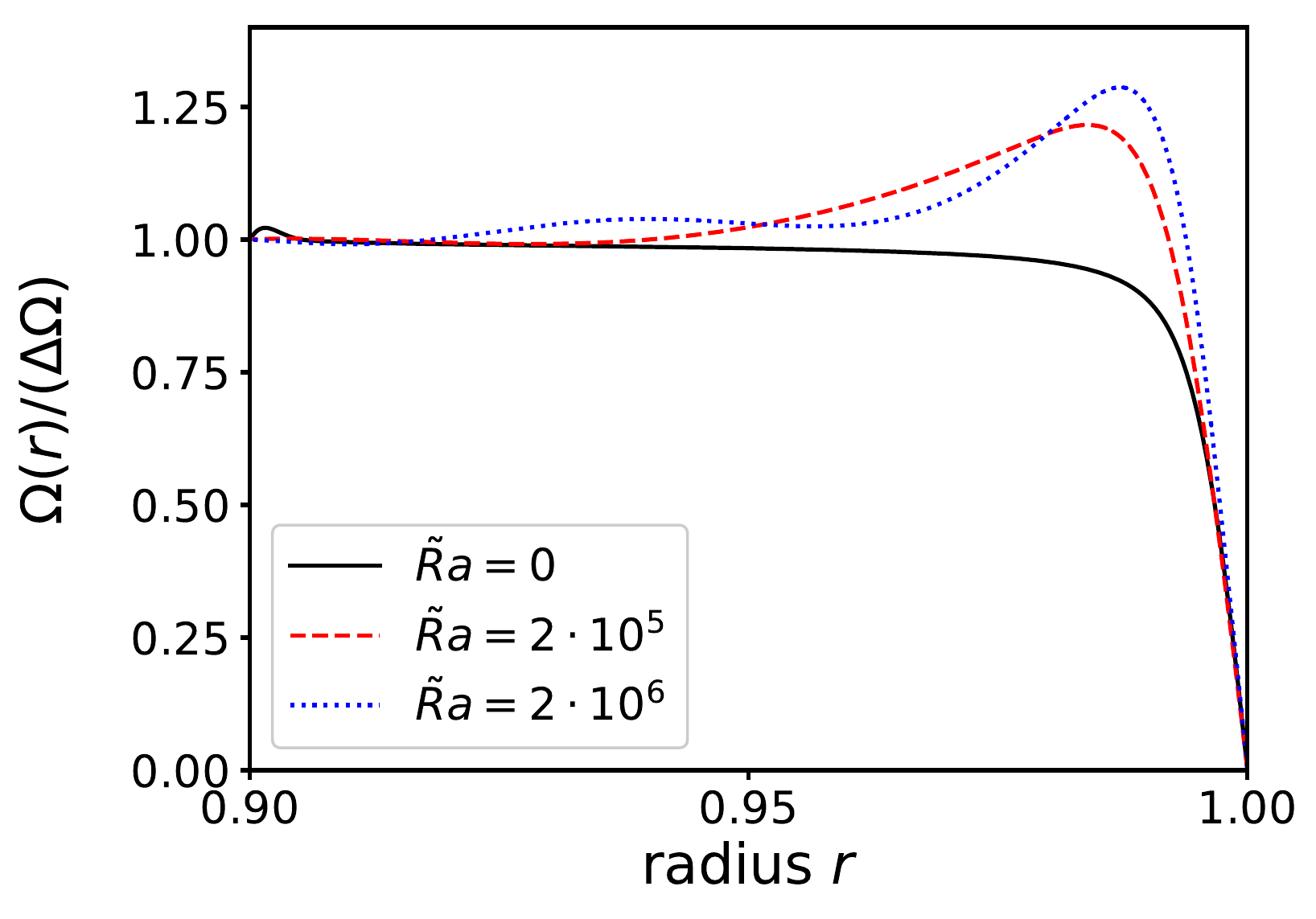}}  \\
\subfigure[$E=10^{-5}$, $\Lambda=1$]{\includegraphics[scale=0.25]{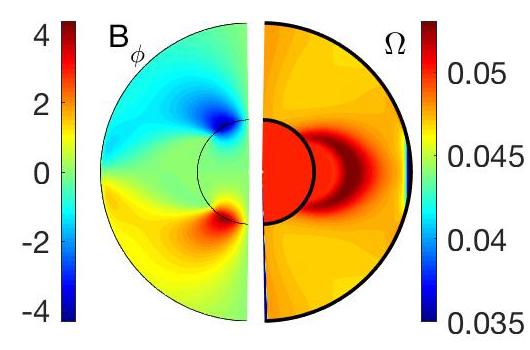}} &
\subfigure[$E=10^{-6}$, $\widetilde{R}a=10^5$]{\includegraphics[scale=0.27]{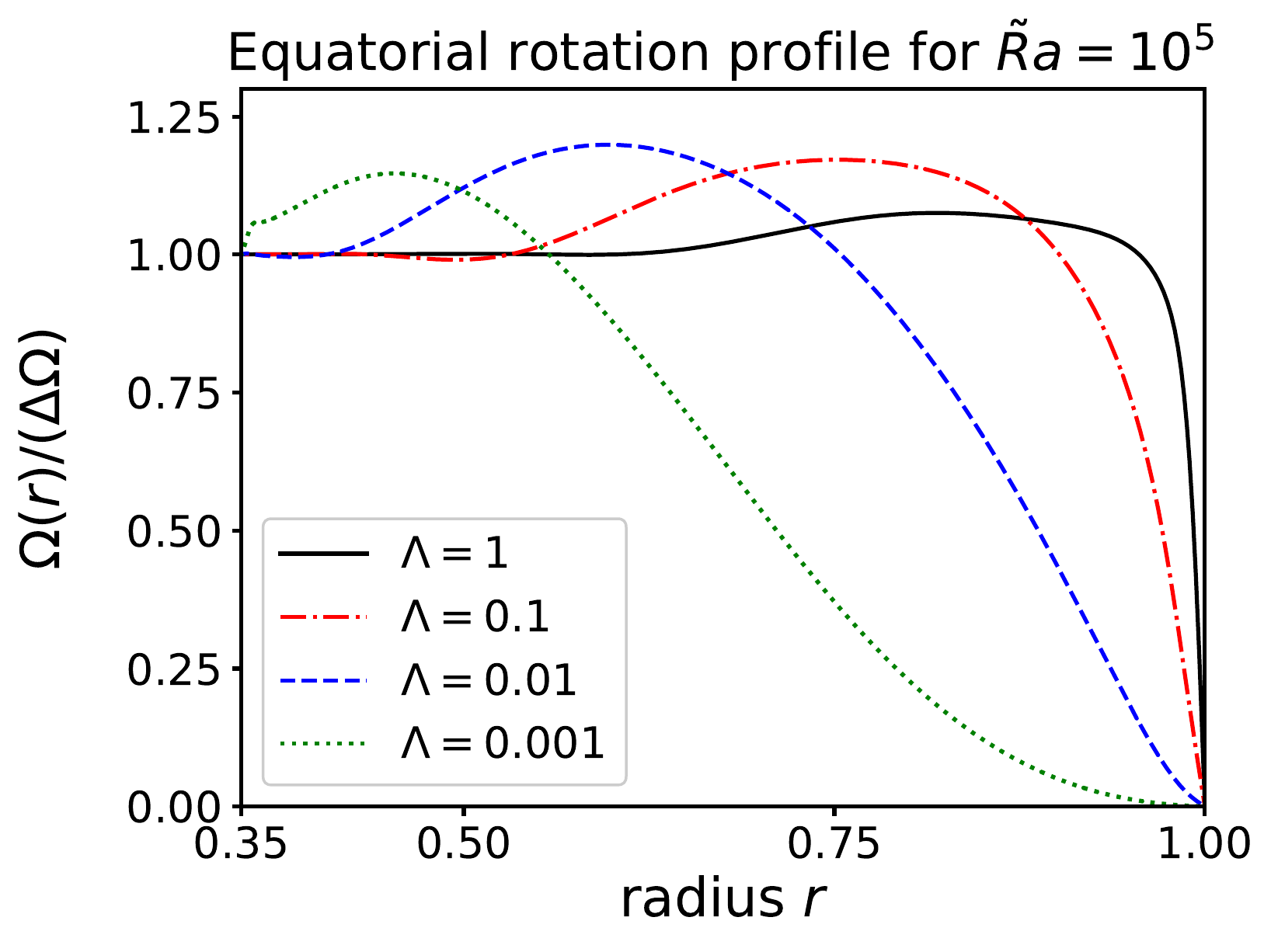}}&
\subfigure[$E=10^{-6}$, $\Lambda=1$, $Pr=0.05$]{\includegraphics[scale=0.27]{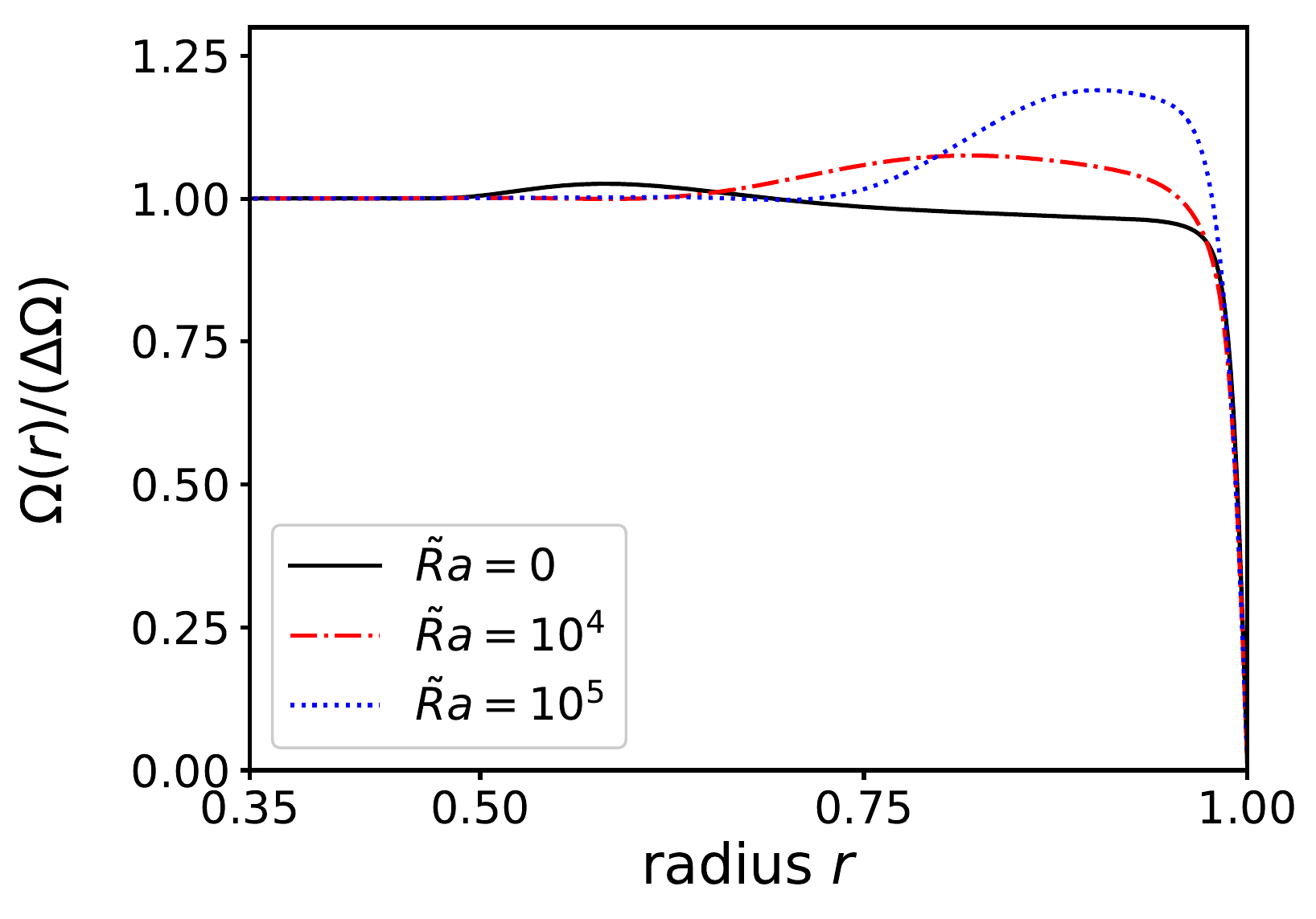}}
\end{tabular}
\caption{On the left (panel $a,d$) : Azimuthal magnetic field $b_\phi$ (left) and rotation profile (right) for $\Lambda = 1$, $Ro = 0.01$, $Pr = 0.05$ and $E = 10^{-5}$, for two different aspect ratios ($r_i/r_o = 0.9$ for the top row, $0.35$ for the bottom row). Otherwise, equatorial rotation profile normalized by the differential rotation $\Delta \Omega$, shown for different values of $\Lambda$  at $E = 10^{-6}$, $\widetilde{Ra} = 2\cdot 10^{5}$, $Pr = 0.05$, (panel (b,e)), and for different values of $\widetilde{Ra}$ at $\Lambda = 1$, $Pr = 1$, $E=10^{-7}$ (panel (c)). Regardless of the aspect ratio (thin gap for the top row and thick gap for the bottom row), super-rotation increases with the stratification ($\widetilde{Ra}$). The color bar of panel (d) has been shifted for visibility purposes (color online).}
\label{fig:super-rotation}
\end{figure*}

\section{Super-rotation induced by the combined influences of magnetism and stratification}\label{sec:superrot}

\begin{figure*}[t!]
\centering
\begin{tabular}{cc}
\includegraphics[scale=0.44]{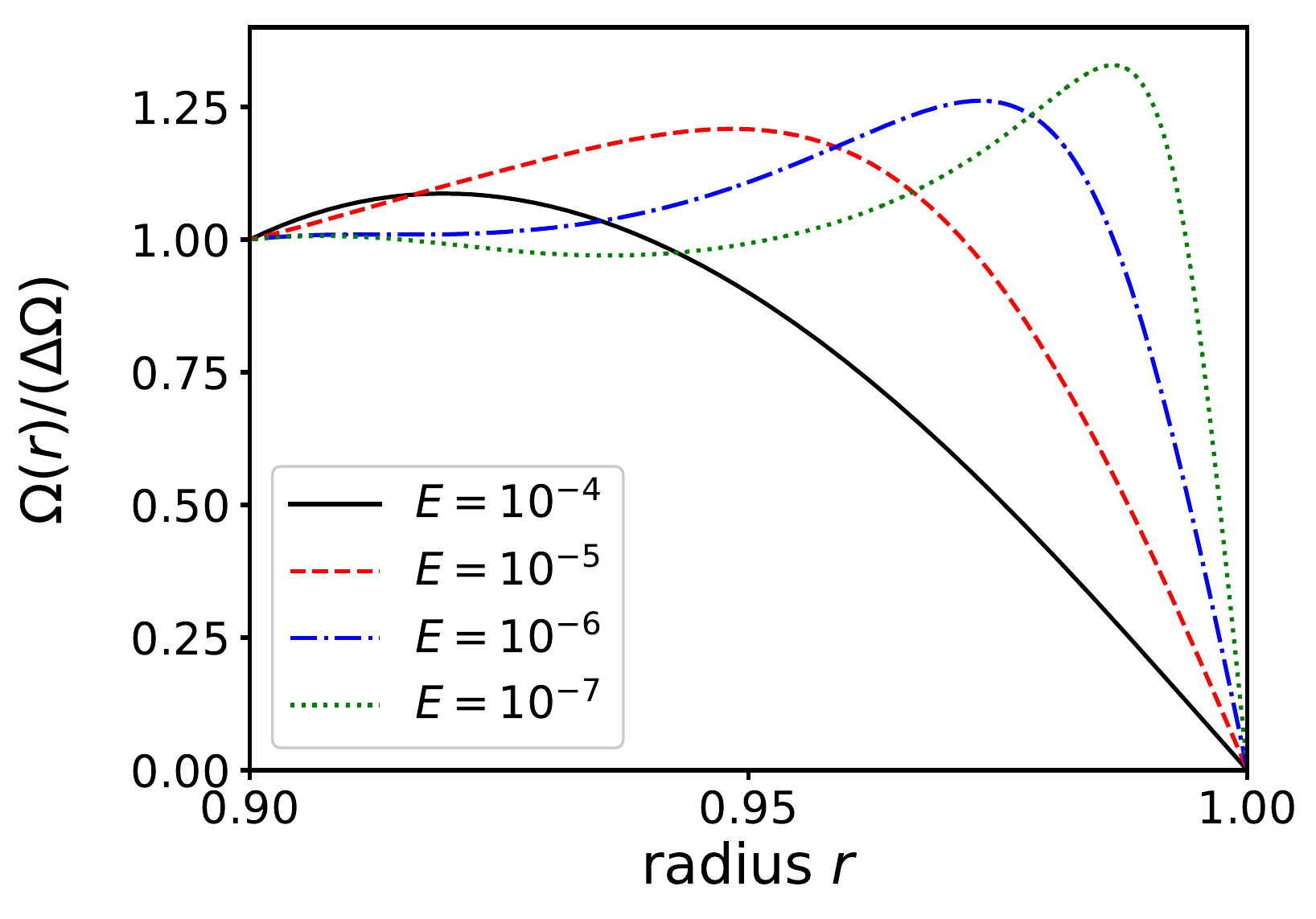} &
\includegraphics[scale=0.35]{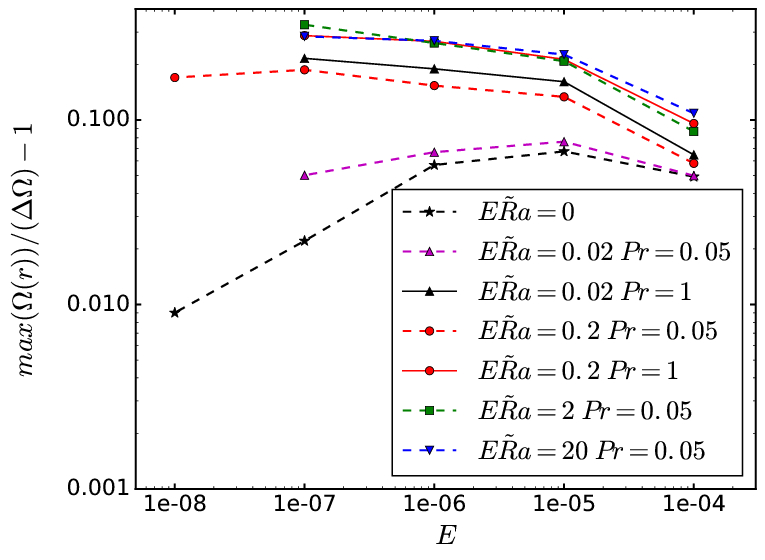} \\
\includegraphics[scale=0.44]{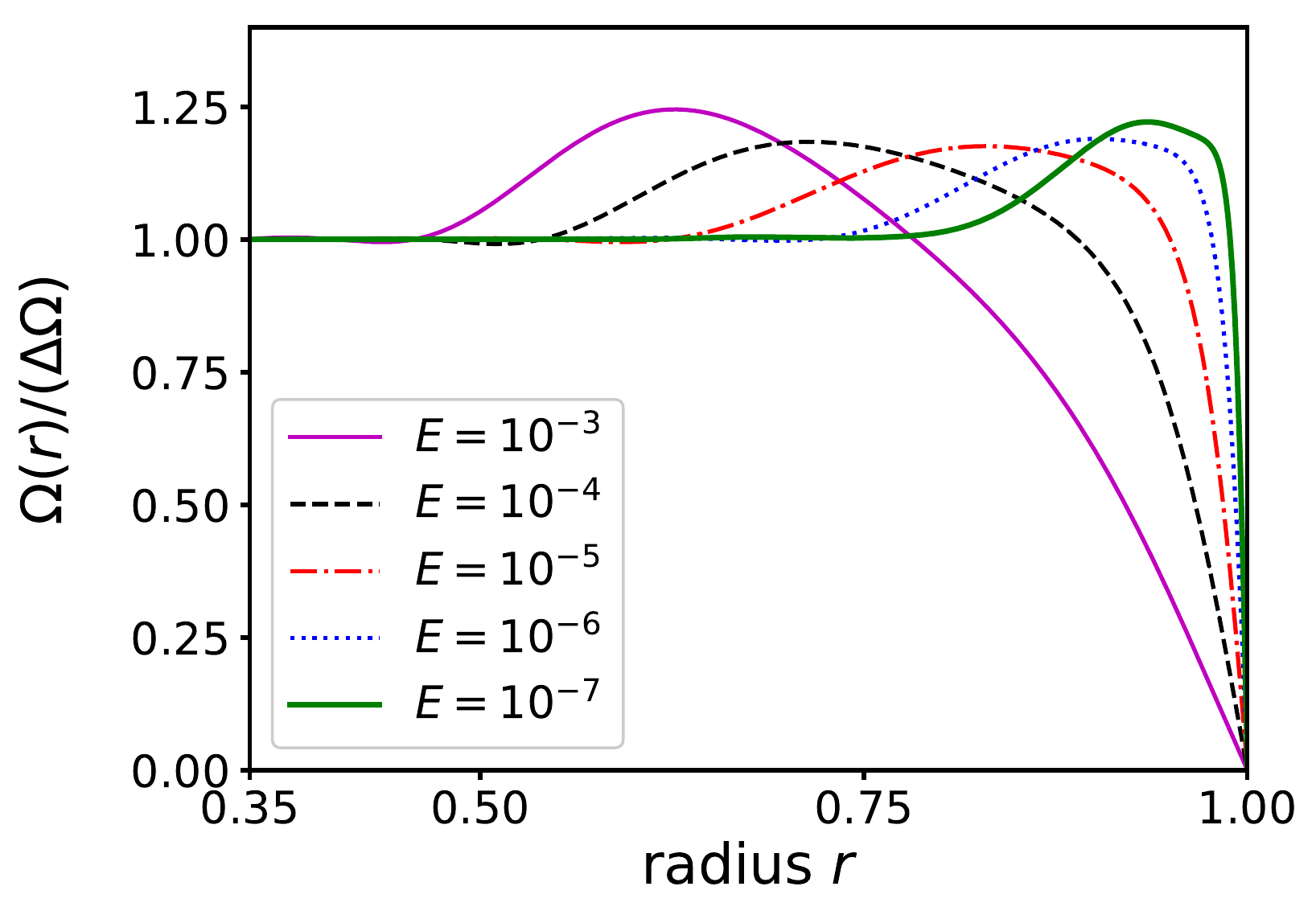} &
\includegraphics[scale=0.35]{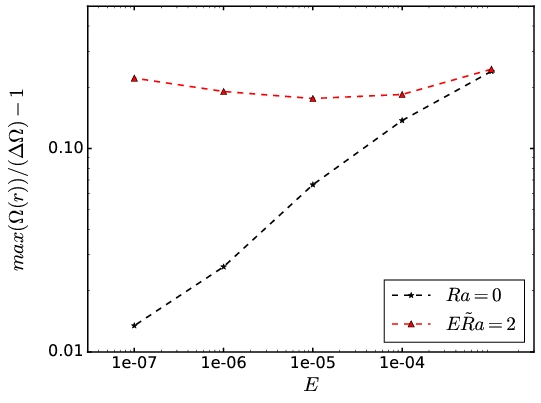}

\end{tabular}
\caption{\textbf{Left}: equatorial rotation profiles for different values of $E$ ($\widetilde{Ra}\times E = 2$, $\Lambda = 1$, all other parameters same as figure \ref{fig:hydro_flows}), showing the generation of a strong shear layer at low Ekman number. \textbf{Right}: Maximum amplitude of the equatorial angular velocity $\Omega_\text{max}$ normalized by the applied differential rotation $\Delta \Omega=\Omega_i-\Omega_o$ for different values of $Pr$ and $E\times \widetilde{Ra}$, all other parameters same as figure \ref{fig:hydro_flows}. Sufficiently large stratification leads to super-rotation independent of $E$ at low Ekman number (color online).}
\label{fig:super-rotation_ekman}
\end{figure*}

We carry out the same direct numerical simulations (DNS) as presented in the previous section, this time adding a dipolar magnetic field. As before, we place ourselves in the low-Rossby regime. We first show in the left panel of figure \ref{fig:super-rotation} the rotation profile $\Omega(r, \theta)$ and corresponding azimuthal magnetic field for a flow where no stratification is applied. Note that the linearity of the equations implies that only the relative values for different regions make sense. It also implies that a change of the Rossby number would only impact the absolute value of the angular velocity $\Omega$, so that we fix it throughout this section at an arbitrary value ($Ro = 0.05$). In total, we performed 120 simulations, where we varied the Prandtl number, the Ekman number and the modified Rayleigh number (according to the caption of the top-right panel of figure \ref{fig:super-rotation_ekman}, where each point corresponds to one simulation), as well as the Elsasser number (according to the caption of figure \ref{fig:super-rotation}(e) ). The aspect ratio was fixed alternatively to $r_i / r_o = 0.9$ and $0.35$.

As illustrated by figure \ref{fig:super-rotation}(a), the structure of the flow is quite different from the hydrodynamic case, in which the rotation rate shows values halfway between that of the inner and outer spheres. Here, the magnetic field lines corotate with the conducting inner core, and the magnetic torque naturally produces a corotation of the majority of the flow with the inner core, except at the outer boundary layer where viscosity effects are no longer negligible, and the no-slip condition makes the fluid corotate with the outer boundary. However, figure \ref{fig:super-rotation}(b) shows that for sufficiently large magnetic field, part of the fluid located in the equatorial region rotates {\it faster} than the inner sphere. This phenomenon is commonly referred to as super-rotation \citep{dormy02} and has been observed in both numerical simulations \citep{hollerbach07} and experiments \citep{nataf06} of magnetized spherical Couette flow. The mechanism involved is the following: since the magnetic field lines corotate with the conducting inner core, the bulk flow will therefore also corotate with it. So does the inner boundary layer, because the current can recirculate inside the conducting inner core there; however, the insulating outer boundary layer can not, which means that the current cannot align with the dipolar magnetic field lines. It is therefore subjected to azimuthal Lorentz forces inducing an extra azimuthal velocity, that can be so high as to overcome the rotation rate of the core. This extra rotation rate propagates to the entire field line, thus explaining the characteristic crescent shape of the super-rotating region.

Unfortunately, this fascinating phenomenon is known to vanish in the presence of a strong global rotation \citep{dormy02}. As shown in the following, it can however be restored in the presence of sufficiently large stable stratification, allowing super-rotation to play a role in the magnetostrophic regime relevant to the planetary context ($\Lambda \sim 1$, $R_o \ll 1$ and $E\ll 1$). figure \ref{fig:super-rotation}(c) shows rotation profiles at the equator for $\Lambda=1$ and different values of the Ekman number $E$. Regardless of the value of $E$, super-rotation increases as $\widetilde{Ra}$ increases and the super-rotating region migrates outward. As a result, we observe the emergence of a localized shear layer close to the CMB when the stratification is sufficiently high. Note that the position of the super-rotating region on the equator remains sensibly the same when $\widetilde{Ra}$ is increased, while its magnitude becomes larger.

Interestingly, the presence of stratification also strongly modifies the behavior of super-rotation in the limit of small Ekman number. Here we take the Earth context as an example: the left panel of figure \ref{fig:super-rotation_ekman} shows the evolution of the rotation profiles with the global rotation for a given value of $Q$ and $\Lambda$, which have both been fixed to their value expected for Earth's outer core, $Q \sim 2$ and $\Lambda \sim 1$. As the Ekman number is decreased, the super-rotating region is enhanced, suggesting a possible magnetostrophic super-rotation close to the CMB.

Precise values of $Pr$ and $\widetilde{Ra}$ in the thin stably stratified region of the Earth's outer core are still difficult to constrain, making very conjectural any prediction on the exact structure of such a shear layer. Still, the right panel of figure \ref{fig:super-rotation_ekman} strengthens the idea that a super-rotating region similar to the one described here may be generated in this layer. Indeed, it shows that for sufficiently large stratification, the magnitude of the super-rotation seems to tend to a constant value in the magnetostrophic regime, independent of both the Ekman number and the non-dimensional parameter $Q$.

These numerical simulations therefore suggest that the stably stratified layer in Earth's outer core might be characterized by a super-rotating region in which the fluid could rotate up to $30 \%$ faster than the inner core in the frame of reference of the outer boundary. Because this mechanism is associated to a strong differential rotation, it provides a simple source for the generation of the MAC waves generally invoked to account for some of the secular variations of the geomagnetic field. Below, we show that in addition to such waves, the stably stratified layer can also be prone to MHD instabilities evolving on time scales similar to Earth's magnetic secular variations.

\section{Local description of unstable MHD modes}\label{sec:local}

As pointed out above, the presence of stratification may strongly enhance the differential rotation, and leads to very localised shear through the generation of so-called super-rotation. The purpose of the present section is precisely to investigate the waves and instabilities prone to be generated in the stably stratified layer whereas previous linear studies have ignored either the effects of differential rotation \citep{buffett16} or the effects of stratification \citep{petitdemange13} in the planetary context. In the stellar context, our stability analysis is strongly inspired by \citet{menou04}. By using recent observational constraints and a simpler linear analysis, we argue on the generation of unstable modes and their possible influences. Note that we only consider axisymmetric modes in the analysis performed in this section.

Because of the force balance achieved between the Lorentz, Coriolis and Archimedes forces, one expects so-called MAC waves to propagate in the stably stratified layer, in addition to Alfvèn waves, purely hydrodynamic waves or torsional oscillations. Furthermore, in the presence of a weak magnetic field, a differentially rotating, conducting fluid can enter an unstable regime and develop what is referred to as the MagnetoRotationnal Instability (MRI). It is essential, for instance, to the understanding of the dynamics of accretion disks owing to the presence of a strong shear in these objects \citep{balbusH98}. Although rotation properties of planetary interiors obviously differ from those of an accretion disk, it was also shown to be impacted by this instability \citep{petitdemangeDB08}. In this section, we study the possibility to see the MRI arise, and especially the impact of stratification and differential rotation on this instability, in two different contexts: planetary CMB and stellar radiative envelopes. In particular, this study stands out in that it extends earlier studies \citep{petitdemangeDB08} to account for the influence of stratification, which had not been accounted for before. Linear calculations show that stratification increases the threshold of the centrifugal instability \citep{richardson20}. However, stable stratification acting on a differentially rotating fluid allows for the development of the so-called StratoRotational Instability (SRI), which amplifies non-axisymmetric modes \citep{molemaker01}. It was studied numerically \citep{shalybkov05} and experimentally \citep{lebars07}. The SRI has no effect on axisymmetric modes.

We follow the linearisation of equation \eqref{eq:full_set} we have used in section \ref{sec:model}, but this time we restrain ourselves to a local analysis and use the WKB approximation to extract a dispersion relation. We place ourselves near the equator where the magnetic field has no radial component. For simplicity, we also do not consider its azimuthal component, which has to match zero at $r = r_o$ as the outer sphere is insulating in our model, so that we are left with the axial component of $\bm{B_0}$ only. Consequently, the state around which we perform the linearisation is described by an azimuthal velocity field $\bm{V_0} =\Omega(s,z) s\bm{e_\phi}$ (with a rotation rate whose gradient we leave free), a uniform vertical magnetic field $B_0\bm{e_z}$, and the stationary temperature $T_s(r)$ described in section \ref{sec:model}. Note that in Earth's case, the thermal, momentum and magnetic diffusivities follow $\kappa \leq \nu \ll \eta$, so that we may drop the thermal diffusivity $\kappa$. We find the following dispersion relation: $a_4\sigma^4 + a_3\sigma^3 + a_2\sigma^2 + a_1\sigma + a_0 = 0$ with
\bse
\label{eq:disp}
\begin{align}
a_4\, &\,=\, {k^2}\big/{k_z^2}\,, \\
a_3\, &\,=\, 2\bigl({k^4}\big/{k_z^2}\bigr)E_\eta + {\widetilde{Ra}E}\big/{k^2} \,,\\
a_2 \,&\,= \,\bigl({k^2}\big/{k_z^2}\bigr)\,(E_\eta^2 k^4 + 2\Lambda_m k_z^2) + 2\widetilde{Ra}EE_\eta - a \,,\\
a_1 \,&\,=\, 2E_\eta\Lambda_m k^4 + E_\eta^2E\widetilde{Ra} k^2 + \bigl({\widetilde{Ra}E}\big/{k^2}\bigr)\Lambda_m k_z^2 - 2aE_\eta k^2\,, \\
a_0\, \,&=\, \Lambda_m^2 k_z^2 k^2 + \widetilde{Ra}EE_\eta\Lambda_m k_z^2 - aE_\eta^2 k^4 - \Lambda_m k_z^2(a+4) \,,
\end{align}
\ese
where
\begin{equation}
a \equiv \dfrac{1}{s^3\Omega^2}\left(\dfrac{k_s}{k_z}\upartial_z(s^4\Omega^2) - \upartial_s(s^4\Omega^2)\right) ,
\end{equation}
and $\Lambda_m = E\Lambda / P_m$, $E_\eta = E/P_m$, $k_s$ and $k_z$ are the non-dimensional radial and vertical component of the wavevector, and $\sigma$ the non-dimensional complex growth rate of the mode of wavevector $\bm{k}$. This dispersion relation is compatible with the one obtained by \citet{menou04}, where both the non-isentropic terms and the thermal diffusivity $\kappa$ are neglected here. While they applied their dispersion relation to the Sun radiative zone, however, the novelty in this paper is that we apply ours to the radiative zone of intermediate-mass stars, as well as to the Earth CMB. Note that, in the case of a purely radial differential rotation (as is the case locally near the equator), the parameter which we denote as $a$ simplifies to $a = -4 + 2Ro'$ where the definition of the shear rate $Ro'$ is based on the angular velocity jump $\delta\Omega$ over a characteristic length scale $\delta s$: $Ro' = (s/\Omega)\delta\Omega / (\delta s)$. We warn the reader that despite the fact they both measure differential rotation, the shear rate $Ro'$ (which is a local quantity) is different from the Rossby number $Ro$ introduced in section \ref{sec:model} (which is a global quantity).

\subsection{Earth's case}

The relevant force balance in Earth's core is the so-called magnetostrophic regime, which allows for the development of a slightly modified version of the classical MRI. We refer to this as the magnetostrophic-MRI, or MS-MRI \citep{petitdemangeDB08}. In particular, the differential rotation and the stable stratification only play minor roles in this force balance. In this regime, a simpler (second order) dispersion relation can be obtained. Note that a similar dispersion relation can also be derived outside the magnetostrophic regime for a stably stratified fluid in the special case where the eddy diffusivities $\eta$ and $\kappa$ are equal:
\begin{equation}
4\Omega^2(\sigma + \eta k^2)^2 + \dfrac{k^2}{k_z^2}(k_zV_{Az})^4 + \alpha g\Delta T(k_zV_{Az})^2 + 2\Omega(k_zV_{Az})^2 s\dfrac{\rmd\Omega}{\rmd s}\,  = \, 0 \, ,
\end{equation}
where $V_{Az}$ is the Alfvèn speed $V_{Az} = B_z/\sqrt{\mu\rho}$. The assumption that the eddy diffusivities are equal relies on the fact that the flow is turbulent. This assumption is not unreasonable, as the flow considered in this section, unlike the previous sections, may be subjected to unstable MRI modes, which may render the flow turbulent. The most unstable mode is purely vertical ($k_s=0$), and its growth rate is:
\begin{equation}
\dfrac{\sigma_r}{\Omega} \, = \, \left(|Ro'| -\widetilde{Ra}E\right)\dfrac{\Lambda/2}{1+\sqrt{1+\Lambda^2}} \, .
\label{eq:strat_is_stable}
\end{equation}
Among the non-dimensional parameters which play a role in the dispersion relation, several are fairly well constrained in Earth's case: we take $E \sim 10^{-15}$, $\Lambda \sim 1$ and $Pm \sim 10^{-5}$ \citep[see][]{book_dormy}. Both differential rotation and stratification however are poorly constrained, and as a result we keep $Ro'$ and $\widetilde{Ra}$ free in the following.

\begin{figure}
\centering
\begin{tabular}{cc}
\includegraphics[scale=0.45]{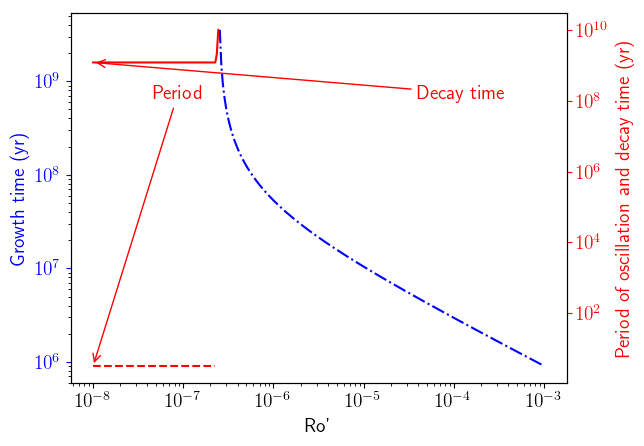} &
\includegraphics[scale=0.45]{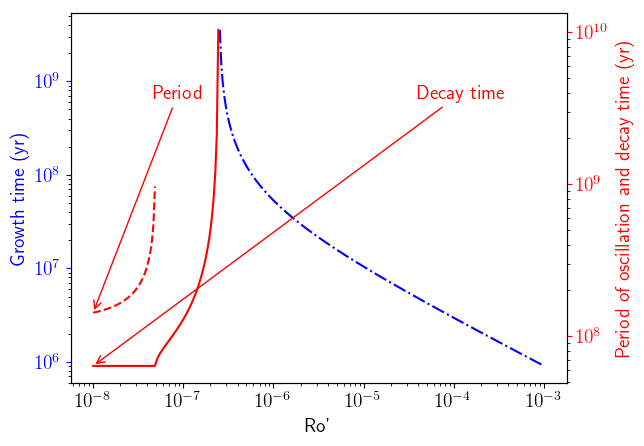}
\end{tabular}
\caption{\textbf{Left :} Evolution of the growth time $T_\text{growth} \equiv 2\pi / \sigma_r$ (\textit{blue dash-dotted line}), decay time $T_\text{decay} \equiv -2\pi / \sigma_r$ (\textit{red solid line}) and period of oscillation $T_\textit{osc} \equiv 2\pi / |\sigma_i|$ (\textit{red dashed line}) of a given wavevector $\bm{k}$ with the shear rate $Ro'$ ($E = 10^{-15}$, $P_m = 10^{-5}$, $\Lambda = 1$ and $\widetilde{R_a} = 10^{10}$; $\sigma_r$ and $\sigma_i$ refer respectively to the real and imaginary part of the complex growth rate $\sigma$). The red part of the plot corresponds to a stable domain for this wavevector, and the blue part corresponds to an unstable domain. \textbf{Right :} Same as left panel, but with $\widetilde{R_a} = 10^{14}$ (color online).}
\label{fig:one_mode}
\end{figure}

The dispersion relation allows us to study the stability of modes characterised by a certain, fixed wavevector $\bm{k}$. In figure \ref{fig:one_mode}, we have selected one wavevector (we chose $\bm{k} = 10\pi\bm{e_z}$ for illustration), and studied both the stability and the oscillatory behaviour of the associated mode when differential rotation and stratification are changed. Note that since $\bm{k}$ is fixed, the modes we describe in figure \ref{fig:one_mode} are in no way the most unstable modes, and do not illustrate the overall stability of the flow. On another note, the dispersion relation has several roots, so that we only selected the one whose growth rate $\sigma$ has the highest real part.

The left panel of figure \ref{fig:one_mode} shows the weak stratification case. For low shear rates, the mode behaves as a damped oscillator, with typical periods much shorter than the damping time, meaning that the wave is very weakly damped. For higher shear rates, however, the wavevector $\bm{k}$ becomes associated to an unstable mode, whose oscillatory part vanishes. This is not surprising, as the differential rotation acts as a reservoir of energy for the MRI.

Adding stable stratification modifies the qualitative behaviour of the modes associated to the wavevector $\bm{k}$. Indeed, one can see from the right panel of figure \ref{fig:one_mode} that it drastically increases the period of oscillation in the low-$Ro'$ regime, to such a degree that it becomes much longer than the decay time. We note the apparition of a third, intermediate-$Ro'$ regime, for which the imaginary part of $\sigma$ actually vanishes, meaning that the mode is no longer oscillating. Thus these figures illustrate what equation \eqref{eq:strat_is_stable} already taught us, that is stable stratification plays a stabilizing role for MS-MRI modes, in the sense that it suppresses stable oscillations and increases the instability growth time.

\begin{figure}
\centering
\begin{tabular}{cc}
\includegraphics[scale=0.38]{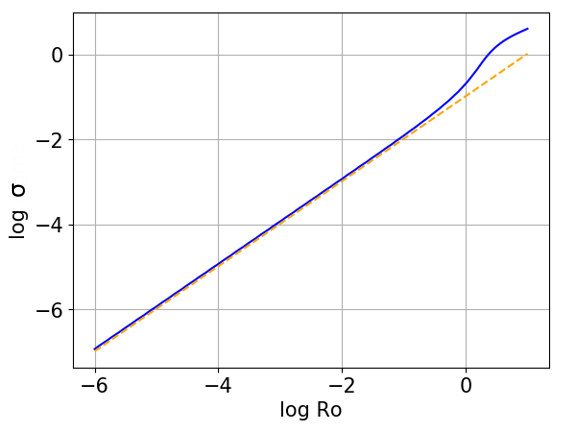} &
\includegraphics[scale=0.38]{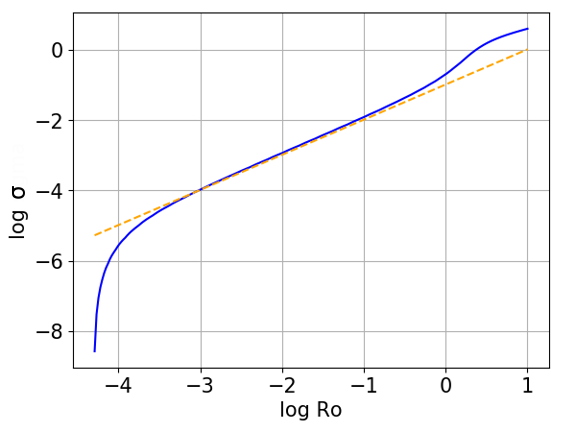}
\end{tabular}
\caption{\textbf{Left :} Evolution of the growth rate $\sigma_r$ of the most unstable mode with the shear rate $Ro'$ ($E = 10^{-6}$, $P_m = 10^{-5}$, $\Lambda = 1$ and $\widetilde{R_a} = 0$). The dashed orange curve shows the analytical solution that can be derived in the magnetostrophic regime (equation \eqref{eq:strat_is_stable}). The solid blue curve shows the solution computed numerically. \textbf{Right :} Same as in the left panel, but with $\widetilde{R_a} = 10^{14}$ (color online).}
\label{fig:all_modes}
\end{figure}

This local study also allows us to search for the most unstable spatial mode for a given set of non-dimensional parameters. Figure \ref{fig:all_modes} shows the evolution of the growth rate of the most unstable mode with the shear rate. In the left panel, we present the non-stratified case: the blue curve shows the numerical computation, with a low-$Ro'$ regime where the real part of $\sigma$ varies linearly with the shear rate, and a high-$Ro'$ regime where it varies with the square-root of the shear rate. We show the prediction from the magnetostrophic regime in orange, which we predictably retrieve for very low shear rates.

We show the stratified regime in the right panel of figure \ref{fig:all_modes}. As we previously mentioned, stable stratification prevents the development of unstable modes at low shear rates. However, it leaves it unscathed when the differential rotation is stronger.

On a final note for Earth's case, we point out that for a shear rate of the order of several percents (which is the typical value obtained if the shear is due to super-rotation), and if we consider that stratification is not high enough to suppress the instability, the numerical study yields $\sigma_r \sim 10^{-3}$ in units of $\Omega_o$ ($\sigma_r$ refers to the real part of $\sigma$), corresponding to a typical growth time $\tau \sim 3$ years, in agreement with the observed typical time scale for the Earth magnetic jerks.

\subsection{The stellar case}

The same dispersion relation can be used regardless of the context. However, the observational constraints are very different in the stellar case and in Earth's case, and we will therefore perform the analysis in the stellar case in a different fashion. In particular, following the stellar community, we prefer to describe the stratification with the Br\"unt-Väisälä frequency $N$ rather than with the modified Rayleigh number, the two being linked \textit{via}
\begin{equation}
\left(\dfrac{N}{\Omega}\right)^2 \,= \,\dfrac{r_o}{H}\widetilde{Ra}E\, .
\end{equation}
where we remind that $r_o$ is the radius of the outer sphere and $H$ the gap between the two spheres in the Couette flow.

In order to choose a parameter range that is relevant to the context of radiative envelopes of massive stars, we make use of the two-dimensional stellar evolution code ESTER \citep{ester}. This code fully takes into account the effect of rotation, and gives us access to realistic values for both the Brünt-Väisälä frequency $N$ and the shear rate $Ro'$. We used a $M = 3~M_\odot$ model, with an equatorial radius $R = 2.4~R_\odot$ and a metallicity $Z = 0.02$. We imposed a surface rotation period of $4.2$ hours, or $70 \%$ of the break-up rotation rate. The radiative envelope of this model yields shear rates that range between $R_o' \sim -0.3$ and $\sim 0.1$; furthermore, if we except the outermost layers of the star, the ratio $N^2 / \Omega^2$ reaches a maximum of $\sim 60$ at $r\sim 0.3~R_\text{tot}$, and stagnates at $\sim 10$ in most of the volume of the radiative envelope. Therefore, we choose to fix $Ro' = 0.1$ in the following, and will vary the order of magnitude of the ratio $N^2 / \Omega^2$ between $1$ and $100$.

\begin{figure}
\centering
\begin{tabular}{cc}
\includegraphics[scale=0.45]{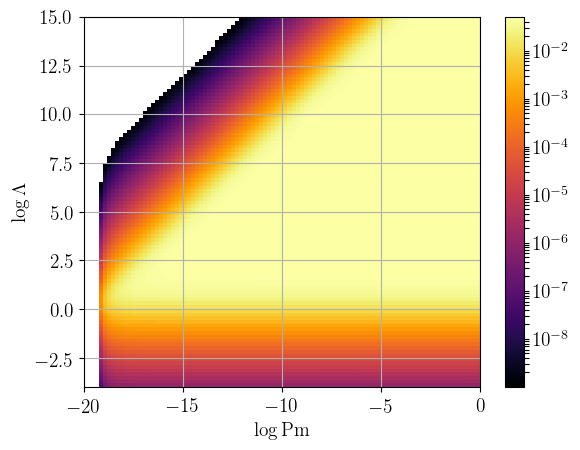} &
\includegraphics[scale=0.45]{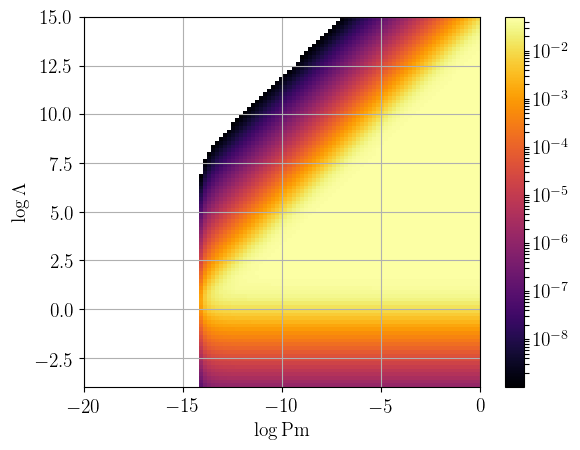} \\
\includegraphics[scale=0.45]{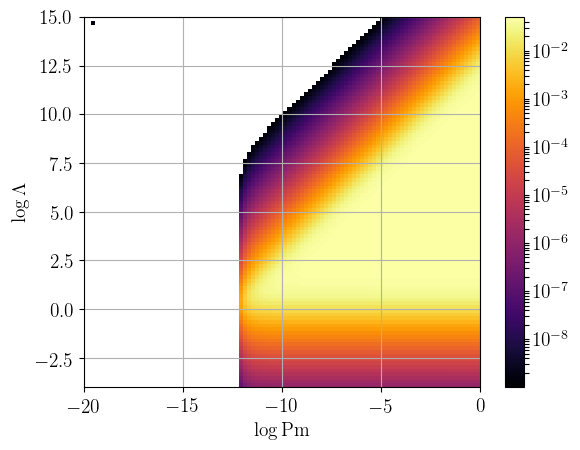} &
\includegraphics[scale=0.45]{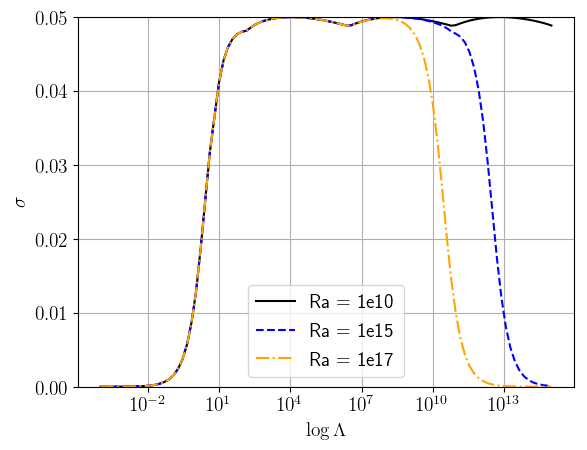}
\end{tabular}
\caption{\textbf{Top left:} Phase diagram of the flow stability in the (log $\Lambda$, log $Pm$) plane, for $E = 10^{-15}$, $Ro' = 0.1$ and $(N/\Omega)^2 = 10^{-5}$ (very weak stratification). The color bar shows the growth rate of the most unstable spatial mode (logarithmic scale) when it exists. The white zones on the left hand side of the phase diagram corresponds to the stable domain. \textbf{Top right:} Same with $(N/\Omega)^2 = 1$. \textbf{Bottom left:} Same with $(N/\Omega)^2 = 100$. \textbf{Bottom right:} Evolution of the real part of $\sigma$ with $\Lambda$, for the same parameters as the three other panels, $Pm = 10^{-2}$ and different values of $\widetilde{Ra}$, showing that the unstable $\Lambda$-window shrinks as the amplitude of stratification is increased. These panels show that the upper bound of the instability window scales as $Pm / \widetilde{Ra}$ (color online).}
\label{fig:stars}
\end{figure}

We present in figure \ref{fig:stars} the results yielded by the dispersion relation \eqref{eq:disp} in this stellar case. The first three panels show the phase diagram of the system stability in the ($Pm$, $\Lambda$) plane for different amplitudes of the stratification. The non-stratified case (top-left panel in figure \ref{fig:stars}) yields a phase diagram that is well-known : there is a $Pm_\text{min}$ below which the instability does not arise, and above that critical value one can see that the system is only unstable if $\Lambda$ is in a certain window, whose width increases with $Pm$. The other panels show that, as we previously encountered with super-rotation, stable stratification has a stabilizing effect on the flow, in the sense that the unstable domain shrinks when $\widetilde{Ra}$ is increased. Note that in order to find the most unstable modes, we have arbitrarily restrained the search to wavectors comprised between $10^{-7}$ and $10^{7}$ in units of the radius gap $H$ between the two shells, for both the radial and vertical components. This was done with the understanding that wavevectors whose norm is below $10^{-7}$ are bigger than the system itself, rendering this local analysis invalid, and that wavevectors whose norm is above $10^{7}$ are killed by dissipative effects (seeing as the magnetic Prandtl number $Pm$ considered here is smaller than one, these dissipative effects would be of ohmic nature). Therefore, while the dispersion relation always formally produce unstable modes regardless of the value taken by the non-dimensional parameters, there is a domain, drawn in white in figure \ref{fig:stars}, in which the only unstable wavevectors are not physically meaningful.

In the last panel of figure \ref{fig:stars} we present a cut of the other panels at fixed x-axis coordinate ($Pm = 10^{-2}$, which corresponds to the value of the magnetic Prandtl number at the top of the Sun radiative core for instance). One can clearly see that the instability window shrinks with increasing stratification. The typical growth rate of the instability within the window remains, however, independent from the stratification.

The results presented above support the idea that, in the parameter range relevant to the context of stellar radiative zones, the MRI is indeed likely to arise, but only if the magnetic field lies in a certain range. Therefore, the MRI may provide a viable mechanism to explain the magnetic desert among intermediate mass stars. Such a dichotomy between stable and unstable magnetic configurations had already been invoked to explain this observational feature \citep[see][]{auriere07,jouveGL15,gaurat15}. However, the results presented in the present paper show that this dichotomy may be caused by an axisymmetric instability, and in particular that the stability of the strong fields harbored by $Ap/Bp$ stars may be due to density stratification.

There is another type of stars in which differential rotation is of primary importance, and those are red giant stars. It is well known, indeed, that these stars feature a strong, localised shear layer near their H-burning shell \citep{deheuvels14}, which is however much weaker than predicted from theory or simulations (the shear rate $Ro'$ inferred from observations is of order unity, while the predicted value would be of order $10^{2-3}$). The MRI is likely to play a crucial role in the dynamics of such region; if that is the case, the saturation of this instability would tend to smooth the differential rotation. Therefore, it is possible that the MRI may play a role in the angular momentum transport needed to account for the observed rotation rate difference between the core and the envelope.

\section{Summary and perspectives}\label{sec:conclusion}

In this study, we have modelled a stably stratified region with an axisymmetric spherical Couette flow embedded in a dipolar magnetic field. We have split this study two ways, by applying this model to two different astrophysical objects: the thin, stably stratified layer close to Earth's CMB; and stellar radiative zones. Note that while we carry out an axisymmetric numerical study, this hypothesis is not necessarily verified in the realistic cases to which we extrapolate our results. We urge the reader to keep the severity of this limitation in mind in the following.

In Earth's context, the most striking feature is the generation of super-rotation in the limit $E\ll1, \Lambda\sim 1$. Combined with MAC waves which are known to transport angular momentum in such a stably stratified layer \citep{buffett16}, the magnetostrophic super-rotation can exert strong shear at the CMB. This new source of shear might be crucial for the understanding of some of the behavior of the Earth's magnetic field, such as the variations in the length of day, or the geomagnetic jerks. Indeed, our local analysis shows that the presence of strong enough shear at the Earth's CMB might trigger MHD instabilities. We showed that the stratification has little direct impact on the growth rate of such instabilities, which arises as long as $\widetilde{Ra}$ remains reasonably low. This suggests that such a flow is likely to be unstable as long as the amplitude $Ro'$ of the differential rotation due to super-rotation remains close to the rough estimate $\sim 10^{-2}$. The stratification enables the presence of such a localised shear by restoring super-rotation (in the magnetostrophic regime) close to the CMB.  In addition, the localisation of super-rotation at the equator may explain why several magnetic features are mainly observed at very low latitude (see for instance \citealt{chulliat14} and \citealt{finlay16}). Furthermore, the observed variations in the length of day might be caused by an acceleration of Earth's mantle due to the non-linear part of the instability. Finally, note that a non-axisymmetric model would be required to explain periodic local magnetic jerks \citep{petitdemange13}. Further investigations on this matter would require a more in-depth study of non-axisymmetric modes in a stratified flow.

In the stellar context, we focussed not so much on the possibility to see super-rotation arise from the interaction between differential rotation and magnetic field (such a shear structure may in this context be subject to large-scale flows, like meridional recirculation, which we have not taken into account) as we have on the potential instabilities which might dramatically affect the flow, in particular the MRI. We show here through a local linear analysis that in the parameter range relevant to the context of stellar radiative zones, the MRI is indeed likely to arise. One of the recognisable characteristic of this instability is that it requires the magnetic field amplitude to lie in a certain instability window. We show that stable stratification plays a stabilising role with respect to the MRI, in the sense that it reduces this unstable window. While it may seem far-fetched to draw quantitative conclusions at this stage, we nonetheless find a dichotomy between stable and unstable magnetic fields, based on their amplitude, that may be a starting point to explain the magnetic desert among intermediate mass stars. The dichotomy based on MHD instabilities has already been proposed in \citet{auriere07} and investigated with 3D and 2D numerical simulations \citep{jouveGL15, gaurat15}. In these studies, the stabilty of strong field configurations is due to the magnetic retroaction on the differential rotation, while here it is an effect of the stable stratification. In any case, whether such dichotomies are at the origin of the observed magnetic desert remains to be proved.

The model we propose here is a simple one, and many aspects of the situation have been overlooked in the process. In particular, we have overly simplified the behavior of the transition region between the convective interior and the radiative envelope at radius $r_i$. While we have imposed a mechanical no-slip boundary condition there, it is well known that, because of inertia, the convective flow in the former overshoots in the latter. The radial velocity entailed by this mechanism thus have on the flow in the stably-stratified layer an impact that may be global; further studies will have to address this point. Furthermore, it is natural that our axisymmetric study be extended to non-axisymmetric, tridimensional simulations, for a lot of instability mechanisms in the studied contexts do not arise in axisymmetric environments. Finally, we have focussed in this paper on the linear part of the instability, before its saturation. While we have strived to draw conclusions about the impact of the saturated phase on the flow, it will be necessary to study it numerically to confirm said conclusions. In particular, only then may the results obtained be quantitatively compared to observations.

Observations are indeed plentiful as far as this study goes. The Kepler mission has in particular given us access to extensive data on giant stars, thanks to which the issue of angular momentum redistribution can be studied in excruciating detail. Furthermore, concerning the intermediate mass stars magnetic desert, new observations of very young stellar objects may help us shed some light on the origin of magnetism in these stars. Tridimensional simulations are also crucial when it comes to understanding the origin of secular variations (SV) in geomagnetism, which are inherently non-axisymmetric. \citet{olsonLR17} have investigated on the matter already, but in a parameter range that did not allow for the development of the MRI. Performing a similar study with much lower Ekman number will allow for the study of the impact of this instability on the geomagnetic SV. The vastness of issues still open on the subject, as well as the innumerable observational data available to help solve them, mean that there is still a lot of work to be done.

\section*{Acknowledgments}

J.P and L.P wish to thank Daniel Reese for giving us access to simulations from the ESTER (Evolution STEllaire en Rotation) code, as well as for his useful comments.

This study was granted access to the HPC resources of MesoPSL financed by the R\'egion \^Ile-de-France and the project EquipMeso (reference ANR-10-EQPX-29-01) of the programme Investissements d'Avenir supervised by the Agence Nationale pour la Recherche. Numerical simulations were also carried out at the CINES Occigen computing centers (GENCI project A001046698). LP acknowledges financial support from ``Programme National de Physique Stellaire'' (PNPS) of CNRS/INSU, France.

\bibliographystyle{gGAF}
\markboth{\rm {J.~PHILIDET ET AL.}}{\rm {GEOPHYSICAL $\&$ ASTROPHYSICAL FLUID DYNAMICS}}

\markboth{\rm {J.~PHILIDET ET AL.}}{\rm {GEOPHYSICAL $\&$ ASTROPHYSICAL FLUID DYNAMICS}}  
\end{document}